\documentclass[pra,twocolumn,letterpaper,floatfix,superscriptaddress,showpacs]{revtex4-1}

\usepackage{graphicx}
\usepackage{amsmath}
\usepackage{braket}
\usepackage{color,soul}
\usepackage{hyperref}
\usepackage{graphicx}
\usepackage[caption=false]{subfig}
\usepackage{graphics}
\usepackage{harpoon}
\definecolor{nblue}{rgb}{0.3,0.3,1.0}
\definecolor{ngreen}{rgb}{0.2,0.7,0.2}
\definecolor{nred}{rgb}{1.0,0.2,0.2}
\definecolor{nyellow}{rgb}{0.9,0.7,0.2}
\definecolor{npurple}{rgb}{0.8,0.2,0.8}
\definecolor{nrose}{rgb}{0.5,0.3,0.3}
\definecolor{nbackground}{rgb}{1,1,1}
\definecolor{ngrey}{rgb}{0.5,0.5,0.5}
\definecolor{nbrown}{rgb}{0.5,0.4,0.0}
\definecolor{nblack}{rgb}{0,0,0}
\definecolor{ncyan}{rgb}{0.1,0.5,0.5}
\definecolor{npale}{rgb}{0.9,0.9,0.9}

\DeclareMathOperator{\E}{E}

\DeclareMathOperator{\ost}{ost}
\newcommand{\dd}{\mathrm{d}}

\newcommand{\grn}{\color{ngreen}}

\newcommand{\blk}{\color{nblack}}

\newcommand{\beq}{\begin{equation}}
\newcommand{\eeq}{\end{equation}}
\newcommand{\nn}{\nonumber}

\newcommand{\erf}[1]{Eq.~(\ref{#1})}

\newcommand{\dg}{^\dagger}

\newcommand{\cu}[1]{\left\{ {#1} \right\}}

\newcommand{\Tr}{\text{Tr}}

\newcommand{\s}[1]{\hat{\sigma}_{#1}}

\newcommand{\past}[1]{\overleftarrow{#1}}
\newcommand{\fut}[1]{\overrightarrow{#1}} 
\newcommand{\both}[1]{\overleftrightarrow{#1}}
\newcommand{\retro}{_{\text R}}
\newcommand{\fil}{_{\text F}}
\newcommand{\sm}{_{\text S}}
\newcommand{\god}{_{\text G}}
\newcommand{\true}{_{\text T}}

\renewcommand{\grn}{\color[rgb]{0.1,0.5,0.1}}
\newcommand{\ti}{ t_0 }

\newcommand{\nbf}{}

\begin{document}

\title{Completely Positive Quantum Trajectories \\ with Applications to Quantum State Smoothing}

\author{Ivonne Guevara}
\affiliation{Centre for Quantum Computation and Communication Technology (Australian Research Council), Centre for Quantum Dynamics, Griffith University, Brisbane, Queensland 4111, Australia}
\author{Howard M. Wiseman}
\affiliation{Centre for Quantum Computation and Communication Technology (Australian Research Council), Centre for Quantum Dynamics, Griffith University, Brisbane, Queensland 4111, Australia}

\date{\today}

\begin{abstract}
To compare quantum estimation theory schemes we must acknowledge that, in some cases, the quantitative difference between them might be small  and hence sensitive to numerical errors.  Here, we are concerned with comparing estimation schemes for the quantum state under continuous measurement (quantum trajectories), namely quantum state filtering and, as introduced by us [Phys. Rev. Lett. {\bf 115}, 180407 (2015)], quantum state smoothing. Unfortunately, the cumulative errors in the most typical simulations of quantum trajectories with time step $\Delta t$ and total simulation time $T$ can scale as $T \Delta t$. Moreover, these errors may correspond to deviations from valid quantum evolution as described by a completely positive map.   Here we introduce a higher-order method that reduces the cumulative errors in the complete positivity of the evolution to of order $T(\Delta t)^2$, whether for linear (unnormalised) or nonlinear (normalised) quantum trajectories. Our method also guarantees that the discrepancy in the average evolution between different detection methods (different `unravellings', such as quantum jumps or quantum diffusion) is similarly small. This equivalence is essential for comparing quantum state filtering to quantum state smoothing, as the latter assumes that all irreversible evolution is unravelled, although the estimator only has direct knowledge of some records. In particular, here we compare, for the first time, the average difference between filtering and smoothing conditioned on an event of which the estimator lacks direct knowledge: a photon detection within a certain time window. We find that the smoothed state is actually {\em less pure}, both before and after the time of the jump. Similarly, the fidelity of the smoothed state with the `true' (maximal knowledge) state is also lower than that of the filtered state before the jump. However, after the jump, the fidelity of the smoothed state is higher.
\end{abstract}
\maketitle

\section{Introduction}

Quantum measurement theory can be considered as the essential link between the quantum properties of a system and the macroscopic world of the measurement apparatuses coupled to it~\cite{W&M10}. The first approach to this matter was introduced by Dirac~\cite{Dir30} in 1930 and almost 30 years later Helstrom  suggested the idea of Positive Operators Valued Measurements~\cite{Hel76} that together with the work of Davies~\cite{Dav76} and Kraus~\cite{Kra83} originated a more generalised version of quantum measurements specially for the case of quantum jumps. Carmichael~\cite{Cha93} suggested the term \emph{quantum trajectory} to describe the stochastic evolution of the quantum states conditioned on the result of continuous  measurements made on the  system's environment,  and `unravellings' for the different types of trajectories arising from different ways of making this measurement. This theory followed a series of individualised quantum measurement experiments~\cite{NSD86, BH&86} and various  related numerical simulations techniques for open quantum systems~\cite{DCM92,GPZ92}.  The same  type of trajectories were also derived from quantum stochastic calculus~\cite{Bel88,B&S92}, and its generalised interpretation in terms of quantum jumps~\cite{Bar90,Bar93}, making the link to the `posterior' conditional state, or `filtered' state from classical estimation theory~\cite{Par81,Jun11}. 

The theory of quantum trajectories {has found many theoretical and experimental applications.} {On the theory side it has been used  in:}  simulations of open quantum systems~\cite{DCM92,GPZ92};  estimation of classical parameters affecting a quantum system~\cite{G&W01,Tsa09a,Tsa09b,Tsa09c, Tsa13,K&M16,Z&M17,SC&16,AR&18,SC&15}  especially adaptive estimation~\cite{Wis95c} in feedback for noise reduction~\cite{Tau95,Buc99,T&W02,Man06} and  in feedback for rapid purification~\cite{C&J06, Jac03,W&B08,CWJ08,LS&13}; and in discovering interesting conditional behaviour for open quantum systems~\cite{K&W11,C&J15,CS&89,CBR91,CC&00}. On the experimental side it has been used in adaptive phase estimation~\cite{AA&02}, conditional state stabilisation by feedback~\cite{SR&02,VM&12}, and analysing typical trajectory behaviour with boundary conditions~\cite{CDJ13}.

A key concern when using quantum trajectories is the robustness of the integration techniques  involved in the numerical simulations.  In naive simulation techniques, the cumulative errors for simulations over a time $T$ using a time step $\Delta t$ can reach a size of order $T \Delta t$, and typically violate positivity. Important work in this area has been done by Kurniawan and James~\cite{K&J05}, and more recently {by} Amini {\it et al.}~\cite{AMR11} and Rouchon and Ralph~\cite{R&R15}, as extensions to the work of Milstein~\cite{Mil95}.  These techniques enable robust simulations which, on average, reproduce the master equation of the system with an error of high order in $\Delta t$. However, they have not explicitly considered in detail the necessity of reproducing completely positive evolution for the trajectories, which is necessary for the trajectories to have an interpretation in terms of quantum measurements on the bath~\cite{W&M10}.  Recent work in \cite{WWC20} shows that the map in   the \blk method in \cite{R&R15} is not completely positive to order $(\Delta t)^2$. 

In this work, we introduce a method to simulate quantum trajectories with a map that is completely positive to order $(\Delta t)^2$, for both quantum jumps and quantum diffusive trajectories. That is, it reduces the cumulative errors in positivity to order O($T (\Delta t)^2$).  Our method is not intended to simulate the quantum master equation to this level of accuracy.  Rather, it is intended to simulate the measurement-conditioned evolution of the quantum state in such a way that the evolution is, to high accuracy, completely positive.  Moreover, when the system is subject to two measurements, if one measurement is held fixed along with its results, while the other measurement is averaged over, the same conditioned evolution for this partially measured system should result regardless of what type of measurement (e.g. homodyne or direct) is implemented for the second measurement.  This ensures that the comparison of different estimation schemes can be done with confidence. These advantages  hold to order O($T (\Delta t)^2$) and rely  on the direct application of quantum maps on the system.

 We developed this method for simulating the smoothed quantum state in the quantum state smoothing formalism introduced by us in Ref.~\cite{G&W15}.  The evaluation of such formalism requires one to compare various estimation schemes --filtering, smoothing, `true' quantum trajectories-- and the difference between them might be small in some cases, making it crucial to verify that the average over a large ensemble of  quantum trajectory simulations of these estimations agree. 
 
In the following we begin with a short description of the quantum trajectories formalism. This includes a summary of the concepts of open quantum systems and {complete positivity} in section \ref{sec:OQS},  including  the relationship between positivity and quantum measurement theory in section \ref{sec:Mop}. This leads on to an introduction to the problem of positivity inherent in the standard quantum trajectory  formalism in section \ref{sec:QTF}. Then, we present our method, which we call completely positive quantum trajectories  (CPQT)  in section \ref{sec:CPQT}. In section \ref{sec:S2SMop} we show how to apply the CPQT  method to a system under two monitoring processes: homodyne detection and photodetection and the  discrepancies in average evolutions are indeed found to be small. This warranty of equivalence in average evolutions is essential for comparing the quantum state filtering and the quantum state smoothing. Lastly, in section \ref{Sec:QSS} we use the method to simulate the dynamics of the system when  only the homodyne phocurrent, not the photodection record,  is known by the estimator.  We show for the first time that, in this case, the  surprising result that the purity of the smoothed quantum state is, on average,  smaller than it is for a quantum filtered state both before and after the time of the jump.

\section{Open Quantum Systems} \label{sec:OQS}
We are concerned with open quantum systems whose unconditioned can be described by a \textit{quantum master equation} in the Lindblad form~\cite{Lin76}:
\begin{equation}
\label{eq:master} 
\dot\rho=\mathcal{L}\rho \equiv -
i[\hat{H},\rho] + \sum_{k=1}^{N} \mathcal{D}[\hat{a}_k] \rho. 
\end{equation}
Here $\hat{H}=\hat{H}^{\dagger}$ is the system Hamiltonian, while $\{\hat a_k\}$ are the Lindblad operators, arising from the coupling of reservoirs, to the system, whose action on an arbitrary state $\rho$ is governed by
\begin{equation} \label{eq:lindbladian}
  \mathcal{D}[\hat a_k]\, \bullet = \hat{a}_k \bullet \hat{a}_k^{\dagger} -
  \frac{1}{2} \{\hat{a}_k^{\dagger}\hat{a}_k, \bullet\} .
\end{equation}
For example, in the particular case of two environments ($b$ and $c$) interacting with the system, the unconditioned state's evolution equation of the system is
\begin{equation}
\label{eq:masterbc} 
\dot\rho= -i [\hat H,\rho] + \hat{b}\,\rho\, \hat{b}^{\dag} +\hat{c}\,\rho \,\hat{c}^{\dag}-\frac{1}{2}\{\hat{b}^{\dag}\hat{b}+\hat{c}^{\dag}\hat{c},\rho\}.  
\end{equation}
In the case where the system  of interest is under observation through the environment, the state of the system is said to be conditioned on the measurement results. The definition and analysis of such  a  state is covered by the quantum trajectories formalism, discussed in Sec.~\ref{sec:QTF}.

\subsection{Measurement operations and positivity} \label{sec:Mop}
The physics of Markovian evolution requires the maps describing the evolution of such systems to not only map positive operators (such as the state matrix) to positive operators, but also to maintain the positivity of the state of the system when  it is  entangled with an arbitrary ancilla. Formally, we require

\begin{enumerate}
\item The map has to be trace-preserving. That is, $\Tr[{\cal M}\rho]=1$ for any normalized state $\rho$.
\item $\cal M$ is a linear map on operators. That is, $${\cal M}\sum_j \wp_j \rho_j=\sum_j \wp_j {\cal M}\rho_j.$$
\item The map is \emph{completely positive}. That is,
\begin{equation}
 \varrho\geq0\;  \Longrightarrow\; 
 (  \mathcal{I}_R\otimes\mathcal{M}_Q)\varrho \geq0;
\end{equation}
where $\mathcal{I}_R$ denotes the identity map on the ancilla system $R$, and $\varrho$ 
 is an arbitrary state on the joint system of $R$ and $Q$. That is, if $\mathcal{M}$ acts only on $Q$, then the extended map $ (\mathcal{I}_R\otimes\mathcal{M}_Q) $ still results in a valid state (up to normalisation). 
\end{enumerate}
It can be proven that any map satisfying the above can be written in an operator sum representation or Kraus representation~\cite{Kra83}  
\begin{equation}
\mathcal{M}\rho=\sum_j \hat{M}_j\,\rho\,\hat{M}_j^{\dagger}
\end{equation}
for some set of operators $\hat{M}_j$ that satisfy the completeness relation,
\begin{equation}
\sum_j \hat{M}_j^{\dagger}\hat{M}_j= \hat{1}.
\end{equation}

 An essential part of any physical simulation is to guarantee that it is physically reasonable. However, it is easy to show that the typical numerical simulations of quantum trajectories schemes  cannot guarantee (in general) complete positivity in the maps $\mathcal{M}$ used to describe the evolution~\cite{N&C00, B&P02}. 
These properties can only be satisfied in the quantum trajectories formalism when considering infinitesimal intervals of time in the measurement process or in the simulation. For example, the completeness relation remains valid for measurement operators only to order $\dd t$. In simulations, where the interval is finite ($\dd t \rightarrow \Delta t$) this may result in non-physical states in the evolution, as we will now show.

\section{Quantum Trajectories Formalism}\label{sec:QTF}
A quantum trajectory can be generally described as the stochastic  path in the state space that the state of an open quantum system follows when it is evolving conditioned on a measurement of its environment.  It can be discontinuous (quantum jumps) or continuous (quantum diffusive trajectories)  depending on the kind of measurements that are conditioning the evolution. An ensemble average of {these} trajectories result in an unconditioned evolution that coincides with the master equation of the state of the system~\cite{W&M10,GPZ92,Cha10,Cha93}.

In terms of estimation theory, quantum trajectories are the normalised version of the filtering equation solutions,  or the equivalent to the Kushner-Stratonovich equation for classical systems. Linear quantum trajectories are the unnormalized version of these paths and relate to the classical Duncan-Mortensen-Zakai equation, where a non-normalized state follows the linear differential equation that averages to the correct state matrix, if the measurement results have a probability distribution that follows a particular ostensible distribution~\cite{W&M10}. 

 In the subsequent subsections we introduce these processes of quantum jumps (corresponding, for example,  to a system whose radiative emissions are subject to photon-counting) or quantum diffusion (corresponding to a similar situation but with homodyne or heterodyne detection).
 
\subsection{Quantum Jumps}
Quantum jumps are the simplest quantum unravelling and take place when the measurement record is discontinuous (e.g. photon-counting). The quantum jumps describing the trajectory are labelled by discrete values  $n_t$   of the measurement results, and its conditioned evolution is described by, the unnormalised conditioned state for this model,   
\begin{equation}\label{eq:rhon}
\tilde{\rho}_{{ n}_{t}}(t+  \Delta t)\equiv
 {\cal M}_{{ n}_{t}}\,\rho(t)  ={\hat{M}_{{ n}_{t}}\rho(t) \hat{M}_{{ n}_{t}}^{\dagger}}.
\end{equation}
Here $\hat{M}_{n_t}$ is the measurement operator satisfying the completeness relation
\begin{equation}
\sum_n\wp_{\rm ost}(n)\hat{M}_{{ n}}^{\dagger}\,\hat{M}_{{ n}}=\hat{1} 
\end{equation} 
where $\wp_{\rm ost}(n_t)$ is an `ostensible' probability distribution for having or not a jump.The ostensible probability distribution is arbitrary, as long as it is fixed for a given ensemble for simulation purposes.

The simplest measurement operators  for quantum jumps, when there is only the type of jump, are
\begin{equation}\label{eq:measopdt}
\begin{array}{rlcrl}
\hat{M}_{ { 0 }} &= 
\hat{1}-\frac{1}{2}\left(\hat{c}^{\dagger}\hat{c}-\lambda\right) \Delta t.  
&&
\hat{M}_{{ 1 }} &= \displaystyle\frac{\grn\hat{c}\blk}{\sqrt{\lambda}}.
\end{array}
\end{equation}
 with ostensible probabilities given by a Bernoulli distribution with
\begin{equation}\label{eq:postn}
\wp_{\rm ost}(n_t:=1)=\lambda \Delta t\;,\;\;\wp_{\rm ost}(n_t:=0)=1-\lambda \Delta t,
\end{equation} 
where $n_t:=1$ indicates a quantum jump in the evolution and $n_t:=0$ none. For this case the completeness relation is given by
\begin{equation}\begin{array}{rl} \label{eq:completejumps}
\sum_{n_t}\wp_{\rm ost}(n_t)\hat{M}_{n_t}^{\dagger}\hat{M}_{n_t}
&= 1 +O((\Delta t)^2)
\end{array}\end{equation} 
 
Note that the quantum trajectories are independent of the ostensible probabilities, in particular of $\lambda$, and the completeness relation is valid in general.  Thus, there are different unnormalised conditioned states that nevertheless average to the correct $\rho$,
\begin{equation}\begin{array}{rl}  \label{rhounconjump}
\rho(t+\Delta t)=&\sum_{n_t}\wp_{\rm ost}(n_t)\hat{M}_{n_t}\rho(t)\hat{M}_{n_t}^{\dagger} \\
=& \rho(t)  
+ {\cal D}[\hat c]\rho(t)\Delta t +O((\Delta t)^2),\blk
\end{array}\end{equation}
with $\mathcal{D}$ defined in Eq.~\eqref{eq:lindbladian}.  For simplicity, we may choose an ostensible probability distribution that factorizes at different times. 

It is more common to use a normalised quantum state defined by 
\begin{equation}\label{eq:normrho}
{\rho}_{{n}_t}(t+\Delta t)=\tilde{\rho}_{{n}_t}(t+\Delta t)\,\frac{\wp_{\rm ost}(n_t)}{\wp_{n_t}}, 
\end{equation}
with actual probability distribution of the results 
\begin{equation}\label{eq:wp}
\wp_{{ n_t}}=\wp_{\rm ost}(n_t) \Tr\left[{\cal M}_{{ n_t}}\,\rho(t)\right].
\end{equation}
For a given initial state, this conditioned state may be  integrated using standard numerical stochastic methods~\cite{W&M10} 

\begin{equation}\begin{array}{rl}\label{eq:deltarhon}
\rho_{{n}_t}(t+\Delta t)=& \displaystyle\Delta n(t)\frac{\hat{M}_1\rho(t)\hat{M}_1^{\dagger}}{\Tr[\hat{M}_1\rho(t)\hat{M}_1^{\dagger}]}\\&+\displaystyle(1-\Delta n(t))\frac{\hat{M}_0\rho(t)\hat{M}_0^{\dagger}}{\Tr[\hat{M}_0\rho(t)\hat{M}_0^{\dagger}]},
\end{array}\end{equation}
with $\Delta n(t)$ obtained from a Bernoulli distribution with probability $\wp_1 =\Tr[\hat{c}^{\dagger}\hat{c}\,\rho(t)]\Delta t$. 

Writing \erf{eq:deltarhon} explicitly in terms of the stochastic increment in the normalised conditioned state, we obtain 
\beq\label{eq:sthon}\begin{array}{rl}
\Delta \rho_{{n}_t}(t)=&-[\hat{H},\rho(t)]\,\Delta t-\mathcal{H}[\frac{1}{2}\hat{c}^{\dagger}\hat{c}]\rho(t)\,\Delta t\\& +\mathcal{G}[\hat{c}]\rho(t)\,\Delta n(t) 
\end{array}\eeq
using the superoperators  
\begin{equation}\begin{array}{rl} \label{defGH}
\mathcal{G}[\hat{c}]\bullet=&\dfrac{\hat{c}\bullet\hat{c}^{ \dagger}}{\Tr[\hat{c}\bullet\hat{c}^{ \dagger}]}-\bullet,\\
\mathcal{H}[\hat{c}]\bullet =& (\hat{c}-\braket{\hat{c}})\bullet + \rm{h.c.},
\end{array}\end{equation}
defined in Ref.~\cite{W&M10}.  From this, the unconditioned state (\ref{rhounconjump}) can be recovered by replacing $\Delta n$ by its expectation value, 
$\wp_1=E[\Delta n(t)]=\Delta t \,\Tr[\hat{c}\rho(t)\hat{c}^{ \dagger}].$
 
\subsection{Quantum Diffusive Trajectories} \label{Sec:3B}
A different way to unravel the master equation is by quantum diffusion. This arises from `dyne' measurements~\cite{W&M10} like homodyne and heterodyne detection. For the former case, the conditioned states follow a continuous but non-differentiable trajectory,
\begin{equation}\label{eq:rhoy}
\tilde{\rho}_{{ y}_{t}}(t+ \Delta t\blk)\equiv
 {\cal M}_{{ y}_{t}}\,\rho(t)  ={\hat{M}_{{ y}_{t}}\rho(t) \hat{M}_{{ y}_{t}}^{\dagger}},
\end{equation}
with $y_t$ a real number. Here the measurement operators are given by
\begin{equation}
 \hat{M}_{{ y}_{t}} = 1   -i \hat{H} \Delta t  -\frac{1}{2} \hat{b}^{\dagger}\hat{b}\, \Delta t + e^{-i \Phi}\hat{b}\, { y}_{t} \Delta t.
\end{equation}
and the ostensible probabilities are Gaussian distributions with zero mean  and variance $1/{\Delta t}$, 
\begin{equation}\label{eq:posty}
\wp_{\rm ost}({y_t})=\sqrt{\frac{\Delta t}{2\pi}}\exp[-{\frac{1}{2}{y_t}^2\Delta t}].
\end{equation} 

Similar to the quantum jumps case,   the completeness relation for these operators is satisfied but only to second order in  $\Delta t$, 
\begin{equation}\label{eq:completeDiffQT}
\int \dd {y}_t\, \wp_{\ost}({{y}_t})\,\hat{M}_{{y}_t}^{\dagger}\,\hat{M}_{{y}_t}=\hat{1} + O((\Delta t)^2).
\end{equation}
The trajectories in this limit are named quantum diffusive trajectories and their unconditioned state,  can be obtained, similarly, from the classical average of the conditioned states,
\begin{equation}\begin{array}{rl} \label{rhouncondifF}
\rho(t+\Delta t)=&\int \dd {{y}}_t\,\wp_{\ost}({{{y}}_t})\,\hat{M}_{{y}_t}\rho (t)\hat{M}_{{y}_t}^{\dagger}\\
=& \rho(t)  -i[\hat{H},\rho(t)]\Delta t   \\&+ {\cal D}[\hat b]\rho(t)\Delta t +O((\Delta t))^2).
\end{array}\end{equation}
Note that this is of the same form as  the one in Eq.~\eqref{rhounconjump}. 

  The actual probability distribution of the measurement results $\wp(y_t)$ can be calculated, using the expression analogous to Eq.~\eqref{eq:wp}. 
  It is found to be a Gaussian with mean $\braket{e^{i \Phi}\hat{b}^{\dagger}+e^{-i \Phi}\hat{b}}$ and variance $1/\Delta t$,
\begin{equation}\label{EulerGauss}
\wp({y_t})=\sqrt{\frac{\Delta t}{2\pi}}\exp\left[-{\frac{1}{2}({y_t-\braket{e^{i \Phi}\hat{b}^{\dagger}+e^{-i \Phi}\hat{b}}})^2\Delta t}\right].
\end{equation}  
   
  The stochastic evolution of the   normalised   conditioned state  in this  case  can be calculated from the above as
\begin{equation}\label{eq:sthoy}\begin{array}{rl}
\Delta\,\rho_{{y}_t}(t)=& -[\hat{H},\rho(t)] \Delta t+ \mathcal{D}[\hat{b}]\rho(t)\Delta t\\&+ \mathcal{H}[e^{i \Phi}\hat{b}]\rho (t)\,\Delta{w}(t),
\end{array}\end{equation}
with $\Delta{w}(t) = {{y}_t}\Delta t -  \braket{e^{i \Phi}\hat{b}^{\dagger}+e^{-i \Phi}\hat{b}} \Delta t $ being Gaussian white noise~\cite{W&M10} and ${\cal H}$ as defined in \erf{defGH}.  This is a standard solution of the stochastic master equation like the one used for quantum jumps, and the unconditioned state 
(\ref{rhouncondifF}) can be obtained by replacing the so-called innovation by its expectation value, $E[\Delta w(t)] = 0$. \blk

\section{Completely Positive Quantum Trajectories}\label{sec:CPQT}

The completeness relations in Eqs.\eqref{eq:completejumps} and \eqref{eq:completeDiffQT} are satisfied only to order $\Delta t$. In theory, this is no a problem. In practice, however, $\Delta t$ is finite and the cumulative errors can easily be non-negligible. This can be understood by considering   that the number of steps is $S = T/\Delta t$ and the cumulative errors are of order $O(S(\Delta t)^2) =O(T\Delta t)$, which grows with $T$. This motivates extending   the simulation methods to higher orders in $\Delta t$.    The CPQT  method extends the quantum trajectory theory to preserve the completely positivity    by reducing   the cumulative errors.  In this section we introduce the method with a {measurement operation} superoperator $\mathcal{M}_r$ extended to order $O((\Delta t)^2)$, for both quantum jumps and quantum diffusion unravellings.  For simplicity, we will ignore the Hamiltonian part of the evolution in this section, and will reintroduce it in Section \ref{sec:S2SMop}. 

\subsection{Quantum Jumps}
We can use the measurement operator for the detection of one photon in Eq.~\eqref{eq:measopdt} but modify the measurement operators for $\hat{M}_0$ to ensure a CPTP map to order $(\Delta t)^2$. Let us consider the completeness condition for the measurement operators $\hat{M}_0$ and $\hat{M}_1$ ,
\begin{equation}\label{eq:CompletQJ}
 \wp_{\rm ost}(n_t:=1)\hat{M}^{\dagger}_1\hat{M}_1 + \wp_{\rm ost}(n_t:=0)\hat{M}^{\dagger}_0\hat{M}_0= \hat{1}.
\end{equation}

From this relation and the definition for the jump measurement operator $\hat{M}_1(\Delta t) $, the non-jump measurement operator would be given by
\begin{equation}
\hat{M}_0(\Delta t)= \sqrt{\frac{\hat{1}-\hat{c}^{\dagger}\hat{c}\, \Delta t}{1-\lambda \Delta t}}.
\end{equation} 
Expanding to order $(\Delta t)^2$, we have
\begin{equation}\begin{array}{rl}
\hat{M}_0(\Delta t)=&\hat{1}-\frac{1}{2}(\hat{c}^{\dagger}\hat{c}-\lambda)(1+\lambda\Delta t)\, \Delta t\\& -\frac{1}{8} (\hat{c}^{\dagger}\hat{c}-\lambda)^2 (\Delta t)^2+ O((\Delta t)^3).
\end{array}\end{equation}
 The reader may wonder why we have expanded the exact expression of the measurement operator $\hat{M}_0$ instead of keeping it exact. We are interested in considering cases where the system may be in a combination of diffusive and jump evolutions, and we show in the following section that the diffusive case imposes restrictions to order $O((\Delta t)^2)$. Thus we decided to limit to this order also  with quantum jumps.  We can show that these expanded measurement operators obey the completeness relation in Eq.~\eqref{eq:CompletQJ} to order $O((\Delta t)^2)$;
\begin{equation}\begin{array}{rl}
\sum_{n_t}\wp_{\rm ost}(n_t)\hat{M}_{n_t}^{\dagger}\hat{M}_{n_t}=&(\hat{1}-\lambda\Delta t)\hat{M}_0^{\dagger}\hat{M}_0+\lambda\Delta t\hat{M}_1^{\dagger}\hat{M}_1 \\
=&\, \hat{1}+ O((\Delta t)^3).
\end{array}\end{equation}

Following the same procedure the evolution of the unconditioned state is given by
\begin{equation}\begin{array}{rl}
\rho(t+\Delta t)=& \sum_{n_t}\wp_{\rm ost}(n_t)\hat{M}_{n_t}^{\dagger}\rho(t)\hat{M}_{n_t} \label{eq:mecpqtjumps} \\
=& \rho(t)+\mathcal{D}[\hat{c}]\rho(t) \Delta t+\frac{1}{4} \mathcal{D}[\hat{c}^{\dagger}\hat{c}]\rho (t)\, (\Delta t)^2\\& + O((\Delta t)^3),
\end{array}\end{equation}
which is an extension to second order of the Lindblad master equation. It partially coincides with the directly extended master equation of Steinbach et al. in~\cite{SGK95},
\begin{equation}\begin{array}{rl} \label{eq:steinbach}
\rho(t+\Delta t)=&  \rho(t)+\mathcal{D}[\hat{c}]\rho(t) \Delta t\\&+\frac{1}{4} \left( 2\mathcal{D}[\hat{c}](\hat{c}\,\rho(t)\,\hat{c}^{\dagger})+\mathcal{D}[\hat{c}^{\dagger}\hat{c}]\rho (t)\right)(\Delta t)^2\\& + O((\Delta t)^3),
\end{array}\end{equation}
but lacks the term $\frac{1}{2}\mathcal{D}[\hat{c}](\hat{c}\,\rho(t)\,\hat{c}^{\dagger})(\Delta t)^2$. This is an interesting coincidence considering that the method in~\cite{SGK95} was shown to be very accurate when integrating the master equation but the authors did not explicitly study the complete positivity of their maps, which is the focus of this paper. 

  The normalised non-linear conditioned state can also be easily calculated with  Eq.~\eqref{eq:deltarhon}. In this case the actual statistics  is  generated in the same way, with $\wp_1=\Tr[\hat{c}^{\dagger}\hat{c}\rho(t)] \Delta t$.  Similarly to previously, replacing $\Delta n$ by this average 
  will then reproduce Eq.~\eqref{eq:mecpqtjumps}.  

\subsection{Quantum Diffusive Trajectories} \label{Sec:4B}
For the quantum diffusive trajectories we propose a generalization of the measurement operators similar to the one for quantum jumps. 
The extended measurement operators for the diffusive case follow from the completeness relation in Eq.~\eqref{eq:completeDiffQT} extended to $O((\Delta t)^2)$. We assume   the ostensible probability to be a Gaussian with zero mean and variance $1/\Delta t$,  as in Sec.~\ref{Sec:3B}.  Then the existing measurement operators from that section (with $\Phi$ set to zero for simplicity), obey the completeness relation 
\begin{equation}\begin{array}{rl}
\int\,\wp_{\rm ost}({y_t}) \hat{M}_{{ y}_{t}}^{\dagger}\hat{M}_{{ y}_{t}}\dd {y_t} 
=& 1 +\frac{1}{4}(\hat{b}^{\dagger}\hat{b})^2(\Delta t)^2,
\end{array}\end{equation}
which deviates from unity at $O((\Delta t)^2)$. \blk 
To cancel the last term in the completeness relation we can introduce the measurement operators for completely positive quantum diffusive trajectories to be
\begin{equation}\label{eq:MytCPQT}
\hat{M}_{y_t}= \hat{1}+\left(e^{-i \Phi}y_t\, \hat{b} -\frac{1}{2}\hat{b}^{\dagger} \hat{b} \right) \Delta t-\frac{1}{8}(\hat{b}^{\dagger} \hat{b})^2 (\Delta t)^2.
\end{equation}
 and the higher order completeness relation is then
\begin{equation}\begin{array}{rl}
\int\wp_{\ost }(y)\hat{M}_y^{\dagger}\hat{M}_y\dd y
=&\hat{1}+O((\Delta t)^3).
\end{array}\end{equation}
In a similar fashion to the previous cases, the equation for the unconditioned state is 
\begin{equation}\begin{array}{rl} \label{eq:mecpqtdiff}
\rho(t+\Delta t)=&\int \dd {{y}}_t\,\wp_{\ost}({{{y}}_t})\,\hat{M}_{{y}_t}\rho (t)\hat{M}_{{y}_t}^{\dagger}\\
=& \rho(t)+\mathcal{D}[\hat{b}]\rho(t) \Delta t+\frac{1}{4} \mathcal{D}[\hat{b}^{\dagger}\hat{b}]\rho (t)\, (\Delta t)^2 \\
&+ O((\Delta t)^3),
\end{array}\end{equation}
coinciding with Eq.~\eqref{eq:mecpqtjumps}.

Contrary to the quantum jumps case, the actual statistics of the measurement results, appropriate for normalised conditioned states, requires some careful consideration. The generalised measurement operators could introduce a new statistics in the measurement results that may not even be Gaussian. Assuming the wrong statistics might end up in obtaining an evolution that  does not coincide with the extended master equation Eq.~\eqref{eq:mecpqtdiff}. Since Gaussian noise is the simplest to generate, we ask the question of whether  
the new statistics of the actual distribution  
\begin{equation}\begin{array}{rl}
\wp (y_t)=& \wp_{\rm ost} (y_t) \Tr [\hat{M}_{y_t} \,{\rho}(t)\, \hat{M}_{y_t}^{\dagger}]
\end{array}\end{equation}
can be approximated by a Gaussian. 

To check this, we calculated the next-highest order corrections, beyond \erf{EulerGauss}, 
to the first four moments,
 $$\mu_m=\E[{y^m}]=\int^{\infty}_{-\infty}y^m\wp(y) \dd y$$  of the measurement results. 
 These are the mean ($\mu=\mu_1$) 
\begin{equation}\label{eq:mu}
\mu=\E[y_t]=-2{\rm Re}\braket{\hat{b}}-{\rm Re}\braket{\hat{b}^2\hat{b}^{\dagger}}\Delta t,
\end{equation}
variance ($\sigma^2=\mu_2-\mu^2$), 
\begin{equation}\label{eq:sigma}\begin{array}{rl}
\sigma^2=&\displaystyle\frac{1}{\Delta t}-2({\braket{\hat{b}^{\dagger}\hat{b}}+2\rm Re}^2\braket{\hat{b}}) 
,
\end{array}\end{equation}
skewness ($\gamma_1=(\mu_3-3\mu\sigma^2-\mu^3)/\sigma^3$)
\begin{equation}\label{eq:gamma1}
\gamma_1=\left(16{\rm Re}^3\braket{\hat{b}}+12\braket{\hat{b}^{\dagger}\hat{b}}{\rm Re}\braket{\hat{b}}-3\braket{\hat{b}^2\hat{b}^{\dagger}}\right)
(\Delta t)^{3/2},
\end{equation}
and kurtosis ($\gamma_2=(\mu_4-4\mu\mu_3+6\mu^2\mu_2-3\mu^4)/\sigma^4$)
\begin{equation}\begin{array}{rll}\label{eq:gamma2}
\small\gamma_2-3=&-24\braket{\hat{b}^{\dagger}\hat{b}}\Delta t .
\end{array}\end{equation}
We then recalculated the master equation using a Gaussian distribution with the same mean and variance of the found distribution. In this case averaging over normalized states the uncoditioned state is 
\begin{align}
\rho_{\rm Gauss}(t+\Delta t) = & \int \dd {{y}}_t\,\wp_{\rm Gauss}({{{y}}_t})\,\frac{\hat{M}_{{y}_t}\rho (t)\hat{M}_{{y}_t}^{\dagger}}{\Tr[\hat{M}_{{y}_t}\rho (t)\hat{M}_{{y}_t}^{\dagger}]}\\
 = & \ \rho(t+\Delta t)+\gamma_1 {\mathcal F}[\hat{b}]\rho(t)(\Delta t)^{3/2} \nn\\
 &+(\gamma_2-3){\mathcal G}[\hat{b}]\rho(t)\Delta t^{2}.
\end{align}
with $\rho(t+\Delta t)$ following Eq.~\eqref{eq:mecpqtdiff},
\begin{equation}\begin{array}{rl}
\mathcal{F}[\hat{b}]\rho(t)=&-4{\rm Re}\braket{\hat{b}}(2{\rm Re}^2\braket{\hat{b}}-\braket{\hat{b}^{\dagger}\hat{b}})\rho(t)\\
&+(4{\rm Re}^2\braket{\hat{b}}-\braket{\hat{b}^{\dagger}\hat{b}})(\rho(t)\hat{b}^{\dagger}+\hat{b}\rho(t))\\
&-2{\rm Re}\braket{\hat{b}}\hat{b}\rho(t)\hat{b}^{\dagger}
 \end{array}
\end{equation}
and
\begin{equation}\begin{array}{rl}
\mathcal{G}[\hat{b}]\rho(t)=&\left(16{\rm Re}^4\braket{\hat{b}}-12{\rm Re}^2\braket{\hat{b}}\braket{\hat{b}^{\dagger}\hat{b}}+\braket{\hat{b}^{\dagger}\hat{b}}^2\right)\rho(t)\\
&-4{\rm Re}\braket{\hat{b}}\left(2{\rm Re}^2\braket{\hat{b}}-\braket{\hat{b}^{\dagger}\hat{b}}\right)(\rho(t)\hat{b}^{\dagger}+\hat{b}\rho(t))\\
&+\left(4{\rm Re}^2\braket{\hat{b}}-\braket{\hat{b}^{\dagger}\hat{b}}\right)\hat{b}\rho(t)\hat{b}^{\dagger}
 \end{array}
\end{equation}
It is important to notice that there is no point in keeping terms of third order or higher in $\Delta t$. Therefore, from Eqs.~\eqref{eq:gamma1} and \eqref{eq:gamma2}, we can discard the terms that involve skewness and kurtosis as not relevant quantities for the completely positive trajectories to $O((\Delta t)^2)$. Once these terms are discarded the master equation in Eq.~\eqref{eq:mecpqtdiff} is recovered and we can be reassured that the Gaussian distribution is sufficient for the simulations. 

There are other higher order simulation methods that are effective computationally like the Euler-Maruyama or the Euler-Milstein. These methods are weakly and strongly convergent to first order, respectively~\cite{AMR11, Mil95} and more recently the authors in~\cite{R&R15} introduced a more efficient method that is an extension of the method in~\cite{AMR11}.   The method in~\cite{R&R15} is also a completely positive map and like ours agrees to first order in $\Delta t$ with the master equation.  As mentioned in the introduction, it has been shown~\cite{WWC20} that the method in \cite{R&R15} is not completely positive to order $(\Delta t)^2$. Similarly, 
there is no reason to think that  the filtered observed record in~\cite{R&R15} has correct statistics to order $(\Delta t)^2$.

\section{Simulation of two simultaneous monitoring processes}\label{sec:S2SMop}

We can use the CPQT for the simulations of two coexisting monitoring processes  on a system  which also undergoes Hamiltonian  evolution.  Consider an open quantum system with two  groups of  output channels ($b$, $c$). An observer monitors the first group, $b$, yielding the measurement record ${\bf  O}$.  A hypothetical observer monitors the second group $c$, yielding a record ${\bf  U}$.  The `true' state of the system $\rho_{\rm T}(t) \equiv \rho_{{\bf  O}_{t},{\bf  U}_{t}}(t)$  is conditioned on both measurement records.  If $\rho_0$ is pure then $\rho_{\rm T}(t)$ will be pure for all times; no extra conditioning could possibly give a better (more pure) state.  

 The Hamiltonian evolution can be included using the correspondent evolution operator $\hat{V}=\exp[-i\hat{H}\Delta t]$,  that acts independently from the stochastic evolution. 
It is important to remark that this evolution is calculated exactly in time, i.e. no approximation in $\Delta t$ has been done on the evolution operator,
\begin{equation}\label{eq:rhoV}
\rho_V=\hat{V}\rho(t)\hat{V}.
\end{equation}
For definiteness and simplicity, we may consider  the observer's single channel with Lindblad operator  ($\hat{b}$) yielding homodyne photocurrent $y_t$ and the hypothetical observer's single channel  with Lindblad operator ($\hat{c}$)  yielding photon count $n_t$, but any combination of monitoring processes can take place. Each one of conditioned evolution can be calculated with a direct map of the state with the correspondent measurement operators.  

 
 In this section we use the technique of generating normalised states with the true probability distribution. Thus the quantum jumps process is simulated from calculating the conditioned states directly with the measurement operators, with the $\Delta n_t$ obtained from a Bernoulli distribution with probability $\wp_1=\Tr[{\hat{c} \rho_V\blk\hat{c}^{\dagger}}]\Delta t$,
\begin{equation}\label{eq:rhony}\begin{array}{rl}
{\rho}_{{n_t}}'=&\displaystyle\Delta n_t\frac{\hat{M}_1\rho_{{V}}\hat{M}_1^{\dagger}}{\Tr[\hat{M}_1\rho_{{V}}\hat{M}_1^{\dagger}]}\\&+\displaystyle(1-\Delta n_t)\frac{\hat{M}_0\rho_{{V}}\hat{M}_0^{\dagger}}{\Tr[\hat{M}_0\rho_{{V}}\hat{M}_0^{\dagger}]},
\end{array}\end{equation}
Likewise, the diffusive process is calculated from a diffusive record $\dd y_t$ drawn from the the actual (not ostensible) distribution. As discussed in Sec.~\ref{Sec:4B} this can be done, to the approximation required, by generated as a random Gaussian variable with mean and variance given by Eqs.~\eqref{eq:mu} and \eqref{eq:sigma} respectively, where expectations $\braket{\bullet}$ are calculated with ${\rho}_{{n_t}}$.  This results in a true normalised pure state 
\begin{equation}\label{eq:multichanelrho}
\rho_{\rm T}(t+\Delta t)= {\rho}_{{y_t},{n_t}}(t+\Delta t)= \frac{\hat{M}_{y_t}{\rho}_{{n_t}}'\hat{M}_{y_t}^{\dagger}}{\Tr[\hat{M}_{y_t}{\rho}_{{n_t}}'\hat{M}_{y_t}^{\dagger}]},
\end{equation}
conditioned  on both records, with $\hat{M}_{y_t}$ following Eq.~\eqref{eq:MytCPQT}. 

We compare the results from the above nonlinear CPQT method with the standard Euler method for quantum trajectories introduced in Sec.~\ref{sec:QTF}, which corresponds to a combination of Eqs.~\eqref{eq:sthon} and \eqref{eq:sthoy} (with the Hamiltonian only counted once), 
\begin{equation}\label{eq:sthony}\begin{array}{rl}
\Delta\rho_{\rm T}(t)=& 
\Delta t\left\{ -{ i}[\hat{H},\rho(t)]  -\mathcal{H}[\frac{1}{2}\hat{c}^{\dagger}\hat{c}]\rho(t) + \mathcal{D}[\hat{b}]\rho(t)\right\} \\& +\mathcal{G}[\hat{c}]\rho(t)\,\Delta n(t)  + \mathcal{H}[\hat{b}]\rho(t) \,\Delta{w}(t). 
\end{array}\end{equation} 
\blk
\subsection{Numerical test}\label{sec:NumT}
To test this part of the simulation we consider  a simple but interesting  open quantum system, the driven damped two-level atom. The Hamiltonian that describes the driving in the interaction frame is:
\begin{equation}\label{eq:Hamiltonian}
\hat{H}=\frac{\Omega}{2}\hat{\sigma}_x,
\end{equation}
In this section we choose $\Omega=3$ for the Rabi frequency, proportional to the amplitude of the driving field. Here time is measured relative to the spontaneous decay rate $\Upsilon=1$. The radiative damping is described by a Lindblad operator is  $\sqrt{\Upsilon}\hat{\sigma}_-$.  We take a fraction $\eta$ of the fluorescence to be observed by homodyne detection, so $\hat b = \sqrt{\Gamma}\s{-}$ with $\Gamma=\Upsilon\eta$. The remainder is registered by photon counts, with Linbdlad operator $\hat c = \sqrt{\gamma}\s{-}$ with $\gamma=\Upsilon(1-\eta)$. Later we will take these jumps to correspond to an unobserved record, as the most natural unravelling to assume for photons absorbed by the environment. 

\begin{figure}[!bt]
\centering
\includegraphics[width=\columnwidth]{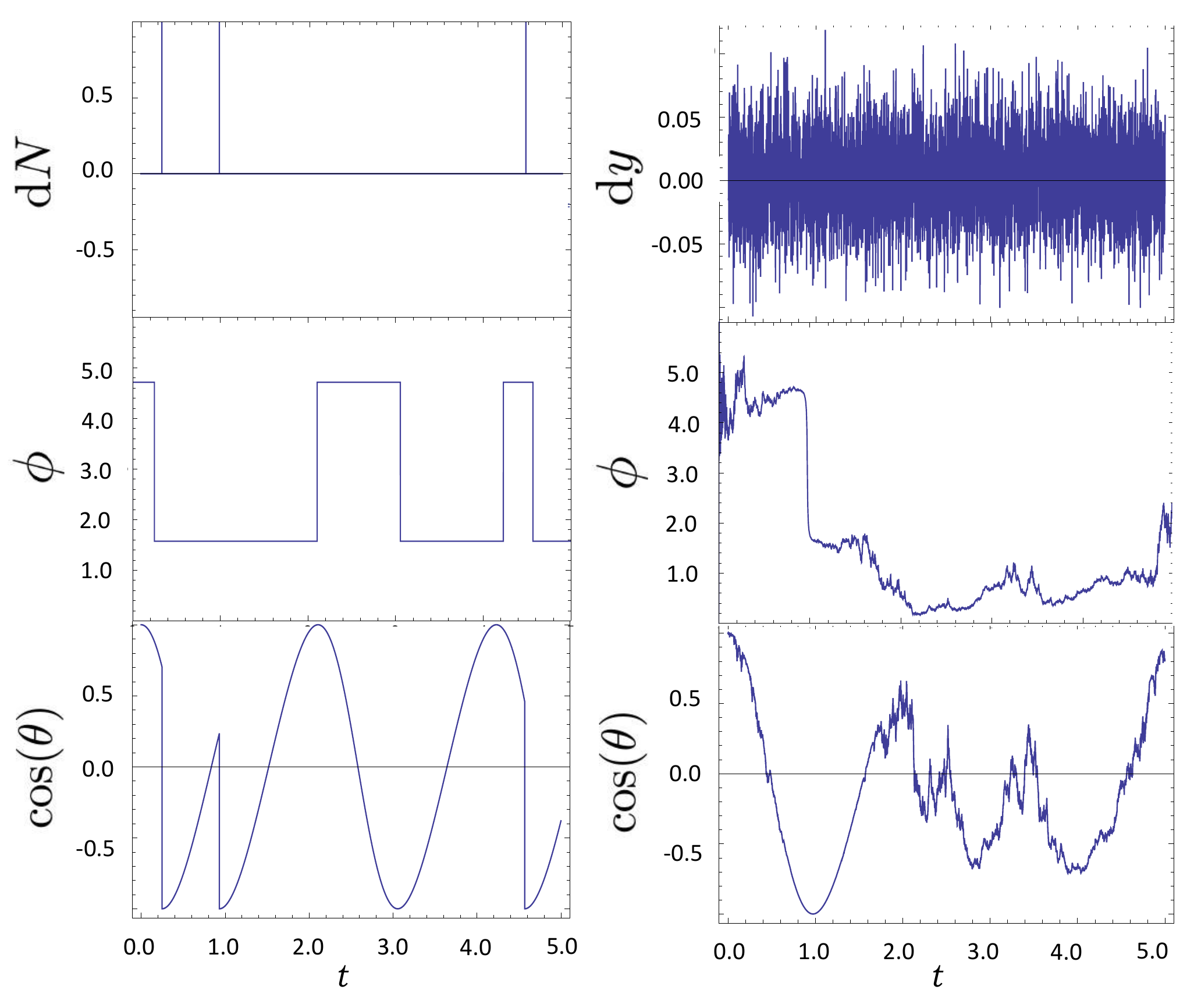}
\caption{[Color online] Measurement records (Top) and segments of a trajectory of duration $5\Upsilon^{-1}$ of an atomic state on the Bloch sphere under photodectection (Left) and homodyne detection with $\Phi=0$ (Right). The  damping, driving, and step size are $\Upsilon=1$, $\Omega=3$, and $\Delta t = 5\times10^{-3}$  respectively. The first and last jumps in $\phi$ on the left correspond to quantum jumps ($dN=1$), but the others correspond to the Bloch vector moving over the north or south pole of the Bloch sphere.}\label{fig:QJdetections}
\end{figure} 

In Fig.~\ref{fig:QJdetections}, we show two typical trajectories on the Bloch sphere, calculated using our nonlinear/normalized/actual-probability CPQT method. The left is photon detection alone ($\eta=0$), while the right is diffusive homodyne detection alone ($\eta=1$). Here $\Phi=0$, corresponding to measuring in quadrature with the mean spontaneously emitted light. The dynamics is as expected and it reflects the typical conditioned evolution for this systems~\cite{W&M10}.

\begin{figure}[!tbp]
\begin{center}
\includegraphics[width=0.9\columnwidth]{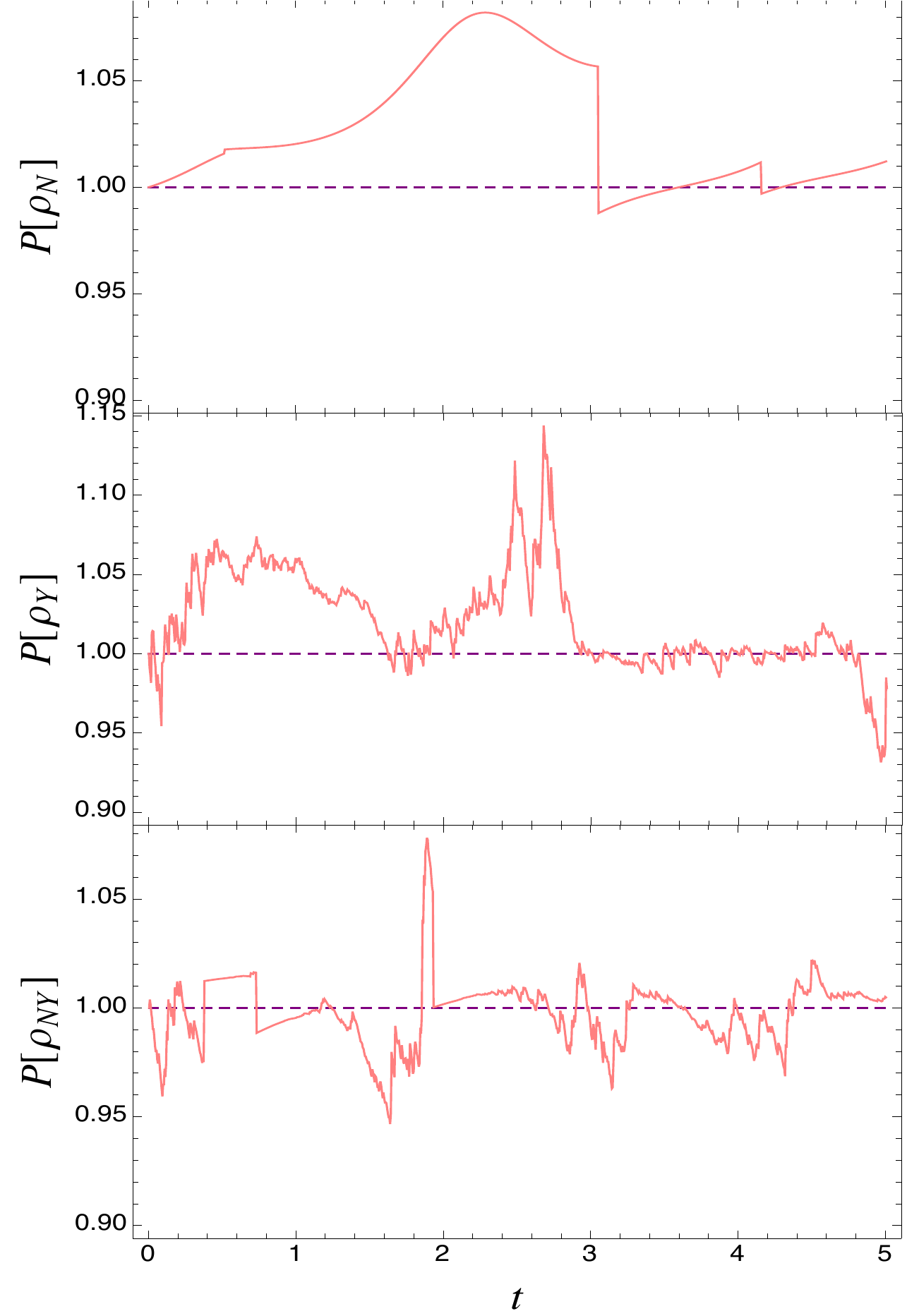}
\caption{[Color online] Comparison of purity of conditioned states between numerical Euler method simulations of quantum trajectories (Solid line) and completely positive quantum trajectories (Dashed line).  The conditioned evolution are Quantum Jumps $\gamma=1$, $\Gamma=0$ (Top), Quantum Diffusive trajectories $\gamma=0$, $\Gamma=1$,  $\Phi = 0$  (Middle)   and multiple channel quantum trajectories $\gamma=\Gamma=0.5$  (Bottom).  The  damping, driving, and step size are $\Upsilon=1$, $\Omega=3$ and $\Delta t = 5\times10^{-3}$.} 
\label{fig:extrapure}
\end{center}
\end{figure}

In Fig.~\ref{fig:extrapure} we compare our CPQT method to the standard Euler method by focussing on the purity of the state, Tr$[\rho_{\rm T}^2]$. We see that in the Euler method the purity often rises above one, which is unphysical. By contrast, using CPQT, the purity remains exactly equal to one for all times,  as it should.

\begin{figure}[!tbp]
  \includegraphics[width=0.9\columnwidth]{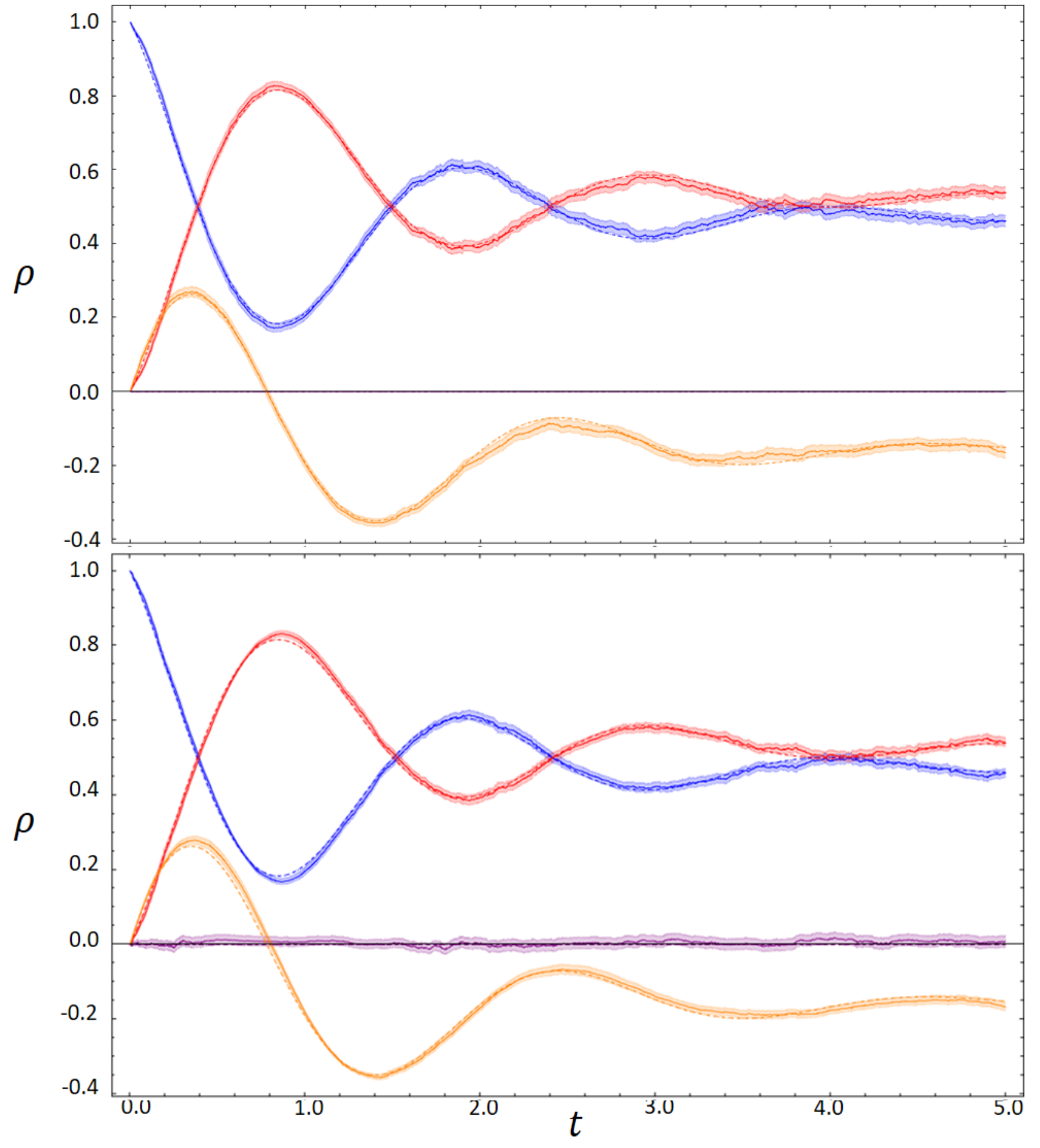}
\caption{[Color online] Comparison of the state components $\rho_{11}$  (blue), $\rho_{00}$ (red), ${\rm Re}[\rho_{12}]$ (purple), and ${\rm Im}[\rho_{12}]$ (orange). The comparison is  between the   unconditioned evolution \blk \unexpanded{$\rho(t)$} (dashed line), the   exact solution of \blk \unexpanded{$\rho_{\rm ME}(t)$} (dot-dashed line) master equation evolution,   and the correspondent ensemble average $E_{{N}_t,{Y}_t}[\rho^{\rm true}_{{N}_t,{Y}_t}(t)]$ (solid line)  over $500$ `true' quantum trajectories from a multiple channel unravelling. Here we have $\gamma=\Gamma$ (that is, $\eta=0.5$) for  photodetection and homodyne detection with local oscillator phase  $\Phi=\pi/2$ (Top) and $\Phi=0$ (Bottom). \blk The  damping, driving and step size are $t=5\Upsilon^{-1}$, $\Upsilon=1$, $\Omega=3$ and $\Delta t = 5\times10^{-3}$.}
\label{fig:CPQTme}
\end{figure}

An  important test on the simulation code is the recovery of the master equation to order $\Delta t$ under ensemble average of a large ensemble of records. In Figure~\ref{fig:CPQTme} the evolution of the components of the matrix state are presented. The evolution of the unconditioned state $\rho_{\rm ME}(t)$ was calculated as the exact solution to the master equation in Eq.~\eqref{eq:masterbc}, obtained with Wolfram Mathematica DSolve. We also solved for the unconditioned evolution   $\rho(t)$ \blk using the higher order unconditional maps of Eqs.~\eqref{eq:mecpqtjumps} and \eqref{eq:mecpqtdiff} and $\hat{V}=\exp[-i\hat{H}\Delta t]$, 
 with $\Delta t = 5\times 10^{-3} \Upsilon^{-1}$. The results from these higher-order maps are indistinguishable from the exact solution of the original master equation. Finally, the plots also present the ensemble average of $500$ CPQTs of the multiple channel monitoring considered. The graphs show how the dynamics predicted analytically 
 are consistent with the exact ensemble average \blk 
\begin{equation}
\rho(t)= E_{{N}_t,{Y}_t}[\rho^{\rm true}_{{N}_t,{Y}_t}(t)]=\rho_{\rm ME}(t).
\end{equation}

In this case, for variety, we also considered a local oscillator phase of $\pi/2$, corresponding to measuring in phase with the mean field from the atom. This has a consequence that the conditioned atomic coherence is completely imaginary, as it is for the master equation solution. \blk 


\section{Application: Quantum State Smoothing} \label{Sec:QSS}
The quantum state smoothing method recently proposed by us in Ref.~\cite{G&W15}, and further explored in Refs.~\cite{LCW20,CGW19,LCW19} is based in the quantum trajectories formalism. This method estimates a positive quantum state that is conditioned on both earlier and later measurement results. Compared with the standard filtering estimation process it can be expected to offer a better approximation to the `true' state that is being estimated. However, the improvement in the estimation can be quite small and we must guarantee the calculations to be as accurate as possible, i.e. the measurement records generated from the true state can be used to construct smoothed states which are comparable to the filtered ones.  Hence it is a priority to ensure that the quantum probabilities are completely valid and independent of the estimation process, as occurs in nature. The CPQT are a fundamental tool to guarantee a fair evaluation of the advantage it offers compared with the quantum filtering estimation. In this section we will show how the completely positive trajectories have been used in Ref.~\cite{G&W15} for this purpose.
 \subsection{General Theory of Quantum State Smoothing}
\begin{figure}[!htbp]
\begin{center}
\includegraphics[scale=.32]{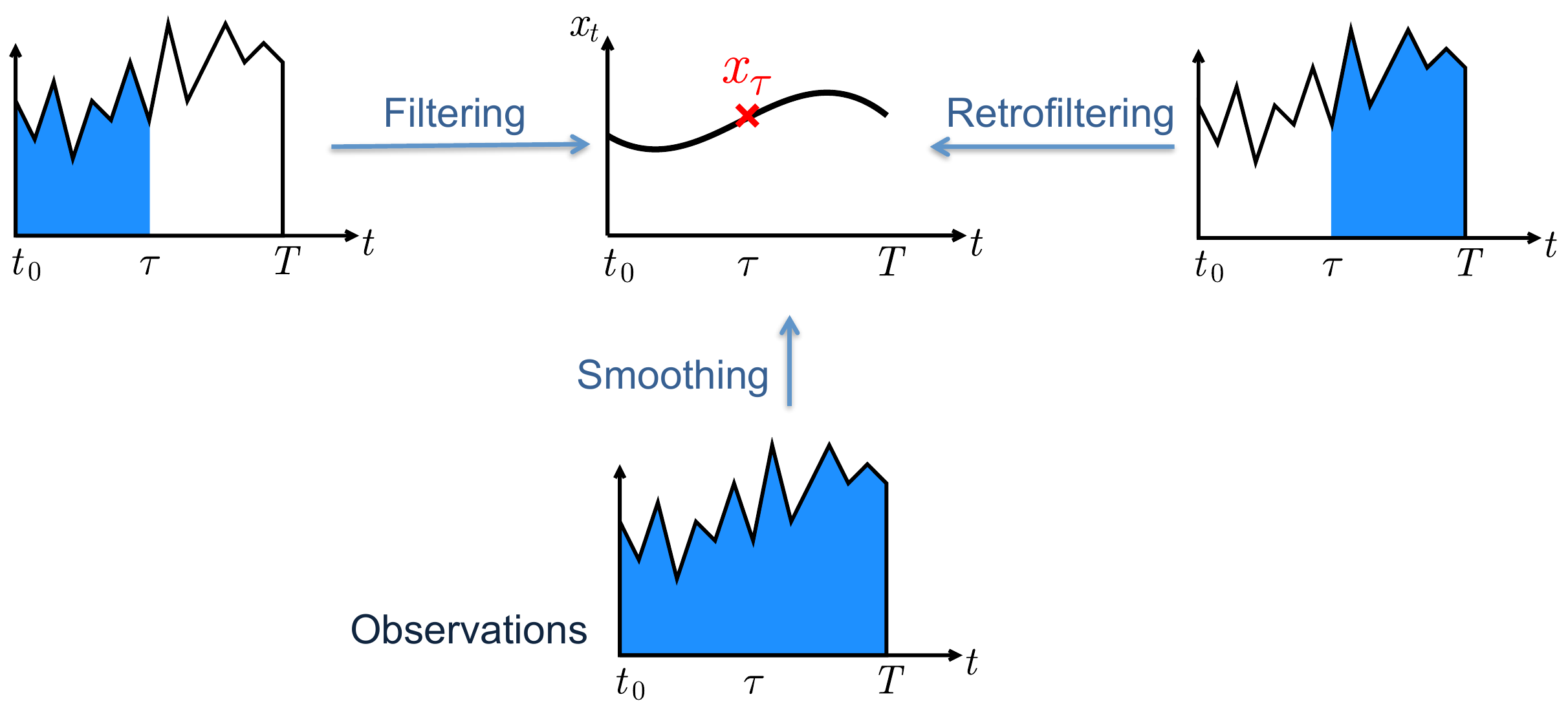}
\caption{[Color online] Classical estimation classes depending on the measurement record considered relative to $\tau$, time at which the signal is to be estimated. (Adapted from~\cite{Tsa09prl}).}
\label{fig:Estimation}
\end{center}
\end{figure}

We will follow the same notation used in Ref.~\cite{G&W15}. We denote a {\em measurement record} ${\bf R}_{\Omega} = \cu{{\bf r}_t\,:  t \in \Omega}$, where $\Omega\subseteq[\ti,T]$ is typically some finite time interval. There are three types of estimation worth distinguishing~\cite{Par81,Jun11}: \textit{filtering}, \textit{retro-filtering} (as we call it), and \textit{smoothing} (see Fig.~\ref{fig:Estimation}). If---as in feedback control problems---for  the time of interest $\tau$ there is only access to earlier results,   $\past{\bf R}_{ \tau} \equiv {\bf R}_{[\ti,\tau)}$, the optimal protocol is filtering. If there is access only to later results, $\fut{\bf R}_{ \tau} \equiv {\bf R}_{[\tau,T)}$,  the optimal protocol is retro-filtering.   As its name implies, this is simply the time-reverse to filtering, but starting with an uninformative final state. Finally, if  the all-time  record $\both{\bf R} \equiv {\bf R}_{[\ti,T)}$ is available, with $\ti<\tau<T$, then all the information can be utilised,  by the technique of  smoothing. 

We have shown in the previous sections that starting with $\rho(t_0)=\rho_0$ and using quantum trajectory theory we can generate the correct filtered probability distribution, while
\begin{equation} \label{eq:and5}
\Tr[\tilde\rho_{\past{\bf R}}\blk
({\tau})] {\wp_{\rm ost}(\past{{\bf R}}_{\tau}|\rho_0)} =  { \wp(\past{\bf R}_{ \tau}|\rho_0)},
\end{equation}
 with 
\begin{equation} \label{eq:rhoFunn}
\tilde\rho_{\past{\bf R}}(t+ \dd t)={\cal M}_{\bf r}\tilde\rho_{\past{\bf R}}(t),
\end{equation}
is the unnormalized state conditioned on the whole past record $\past{\bf R}_{\tau}$. 

The corresponding analogue for Bayesian state retrofiltering has been  set out in~\cite{Tsa09prl};  it is the solution of the adjoint of equation \eqref{eq:filtensav}, 
\begin{equation}\label{eq:Eretro}
\hat{E}\retro(t)\equiv \hat E_{\fut{\bf R}_{t}}(t) 
={\cal M}_{{\bf r}_{t}}^{\dagger}\, \hat{E}\retro(t+\dd t),
\end{equation}
known as the \textit{effect operator}. It evolves backwards from the final uninformative effect $\hat{E}\retro(T)=I$ towards $\hat{E}\retro(\tau)$ conditioned on the record $\fut{\bf R}_{t}$ in the future of $\tau$,  \blk and determines the statistics of $\fut{\bf R}_{\tau}$: 
\begin{equation}\label{eq:TrEyrho}
\Tr[\hat E\retro
(\tau){\rho}_\tau]\wp_{\rm ost}(\fut{\bf R}_{\tau}|\rho_\tau) = \wp(\fut{\bf R}_{\tau}|\rho_\tau).
\end{equation}

\begin{figure}[!btp]
\begin{center}
\includegraphics[scale=.3]{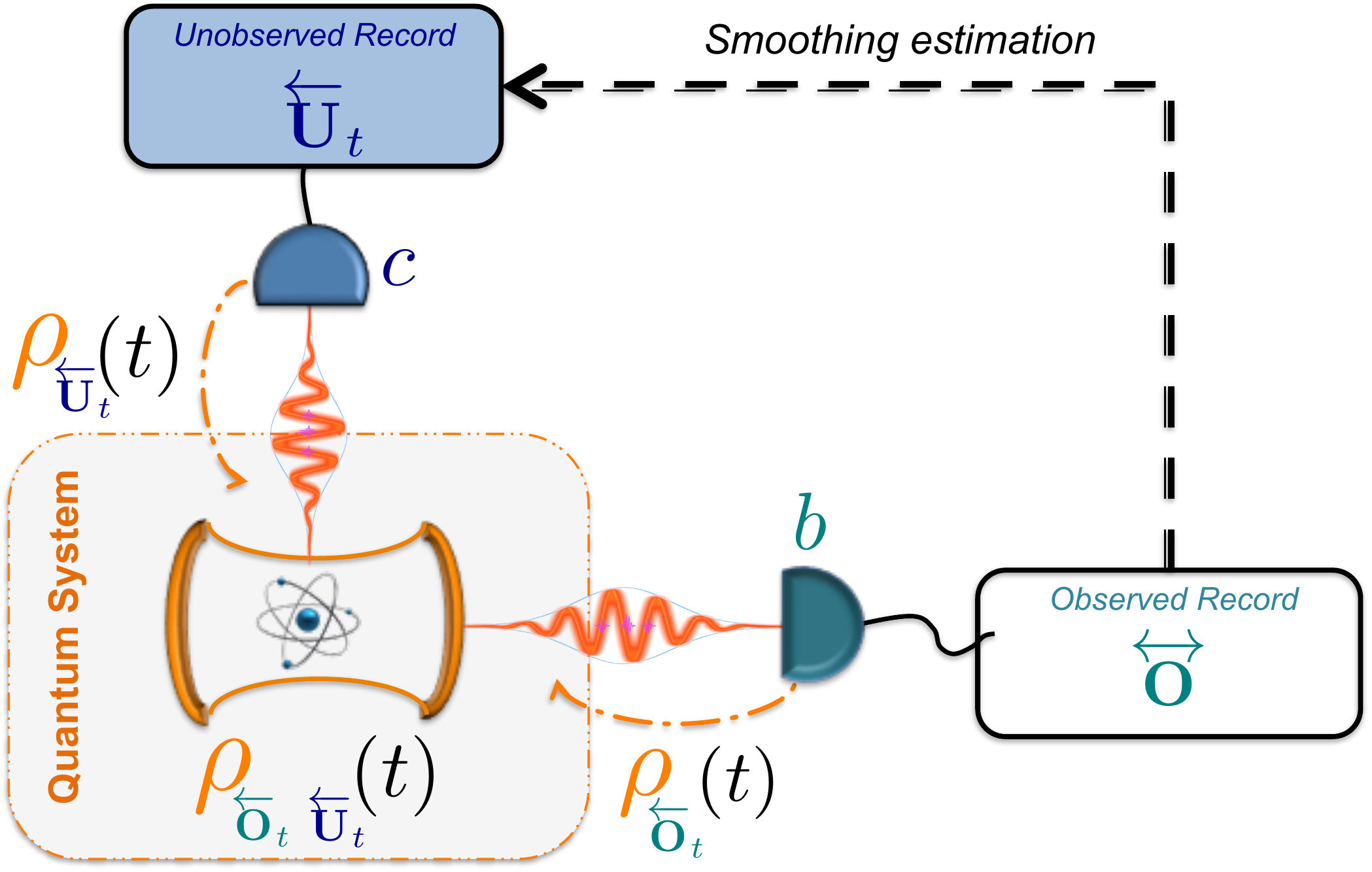}
\caption{[Color online] The quantum state smoothing problem:  to best approximate the unknown true state of a quantum system, conditioned on both observed ($\bf O$) and unobserved ($\bf U$) records, given access only to  $\bf O$. This requires one to estimate \unexpanded{$\past{\bf U}_t$}  up to time $t$ using the full record for  \unexpanded{$\both{\bf O}$}  (before and after $t$).}
\label{fig:QT}
\end{center}
\end{figure} 

To define the quantum smoothed state, we consider the situation, illustrated in Fig.~\ref{fig:QT}, in which an open quantum system (partially observed by experimenter Alice) is coupled to several baths (all assumed Markovian). The experimenter can only monitor a few of the channels, yielding to the observed record $\bf O$. The other baths ---not monitored by her--- could, hypothetically, be monitored by another party, say Bob, yielding results $\bf U$  unobserved by Alice. Under these conditions the `true' state $\rho_{\rm T}(t) \equiv \rho_{\past{\bf O}_{t},\past{\bf U}_{t}}(t)$ conditioned on both Alice's and Bob's records would be pure while  Alice's state, calculated in the conventional way (filtering), 
\beq\label{eq:filtensav}
\rho\fil(t) \equiv \rho_{\past{\bf O}_{t}}(t) = {\rm E}_{\past{\bf U}_{t}|\past{\bf O}_t} [\rho_{\past{\bf O}_{t},\past{\bf U}_{t}}(t) ],
\eeq
conditioned only on $\past{\bf O}_{t}$ would be mixed.  Here ${\rm E}_{A|B}[X]$ means the expected value of $X$, averaged over $A$, for a given $B$.
 In Ref.~\cite{G&W15} we proposed a way for Alice to do better using information in the future of $t$, to learn about the unobserved record. 
The natural generalization of \erf{eq:filtensav} is the smoothed quantum state for time $t$, 
\begin{equation}\label{eq:defsqs}
\rho\sm(t) =  {\rm E}_{\past{\bf U}_{t}|\both{\bf O}} [\rho_{\past{\bf O}_{t},\past{\bf U}_{t}}(t) ] \equiv \sum_{\past{\bf U}_{t}} \wp\sm(\past{\bf U}_{t})  \rho_{\past{\bf O}_{t},\past{\bf U}_{t}}(t), 
\end{equation}
which is by construction also positive-definite. 
Here $\wp\sm(\past{\bf U}_{t})= \wp_{\both{\bf O}}(\past{\bf U}_{t}) 
=\Pr[\past{\bf U}_{t}^{\rm true}=\past{\bf U}_{t}|\both{\bf O},\rho_0]$ is the probability distribution for the unobserved record prior to $t$, obtained by smoothing from  the all-time observed record $\both{\bf O}$.   Elementary manipulation of probabilities~\cite{G&W15} gives $\wp\sm(\past{\bf U}_{t}) \equiv  
\wp(\past{\bf U}_t|\both{\bf O})
\propto \wp(\fut{\bf O}_{t}|\past{\bf U}_{t},\past{\bf O}_{t})\,\wp(\past{\bf U}_{t}|\past{\bf O}_{t})$. 
Using the  equations for multiple channels corresponding to  Eq.~\eqref{eq:TrEyrho},
\begin{equation}\begin{array}{rl}\label{eq:py|ny}
\wp(\fut{\bf O}_{t}|\past{\bf U}_{t},\past{\bf O}_{t})&=\Tr[\hat{E}_{\fut{\bf O}_{t}}\rho_{\past{\bf U}_{t}\past{\bf O}_{t}}]\,\wp_{\text{ost}}(\fut{\bf O}_{t}), 
\end{array}\end{equation}
and to  \erf{eq:and5},  
\begin{equation}\label{eq:TrErhoun}
\Tr[\hat{E}_{\fut{{\bf  O}_{t}}}\tilde{\rho}_{\past{{\bf  U}_{t}}\past{{\bf  O}_{t}}}]\,\wp_{\text{ost}}(\past{{\bf  U}_{t}}|\past{{\bf  O}_{t}	})=\Tr[\hat{E}_{\fut{{\bf  O}_{t}}}\rho_{\past{{\bf  U}_{t}}\past{{\bf  O}_{t}}}]\,\wp(\past{{\bf  U}_{t}}|\past{{\bf  O}_{t}}), 
\end{equation}
we finally obtain, from \erf{eq:defsqs}, 
\begin{equation} \label{eq:tildap}
\rho\sm(t) \propto \sum_{\past{\bf  U}_{t}}\wp_{\text{ost}}(\past{\bf  U}_{t}|\past{\bf  O}_{t})
\times 
{\rho}_{\past{\bf  U}_{t},\past{\bf  O}_{t}}(t) \,\Tr[\hat{E}_{\fut{\bf  O}_{t}}(t) \tilde{\rho}_{\past{\bf  O}_{t},\past{\bf  U}_{t}}(t)].  \\
\end{equation} 
This is the method we use below to find the smoothed quantum state. 

Note that $\wp_{\text{ost}}(\past{\bf  U}_{t}|\past{\bf  O}_{t})$ is the conditional probability distribution calculated as if the ostensible distribution $\wp_{\text{ost}}(\past{\bf  U}_{t},\past{\bf  O}_{t})$ were the true probability distribution. The latter is the distribution for which the $\tilde{\rho}_{\past{\bf  O}_{t},\past{\bf  U}_{t}}(t)$ appearing in Eq.~(\ref{eq:tildap}) is the appropriate unnormalized state. For simplicity, we take it to to be $\wp_{\text{ost}}(\past{\bf  U}_{t},\past{\bf  O}_{t}) = \wp_{\text{ost}}(\past{\bf  U}_{t})\wp_{\rm ost}(\past{\bf  O}_{t})$, so that $\wp_{\text{ost}}(\past{\bf  U}_{t}|\past{\bf  O}_{t})=\wp_{\text{ost}}(\past{\bf  U}_{t})$.  For the ostensible probabilities for the individual records, we use Eq.~\eqref{eq:postn} for photocurrent record, and Eq.~\eqref{eq:posty} for a homodyne photocurrent record. 

\begin{figure}[!htbp]
\begin{center}
\includegraphics[width=0.8\columnwidth]{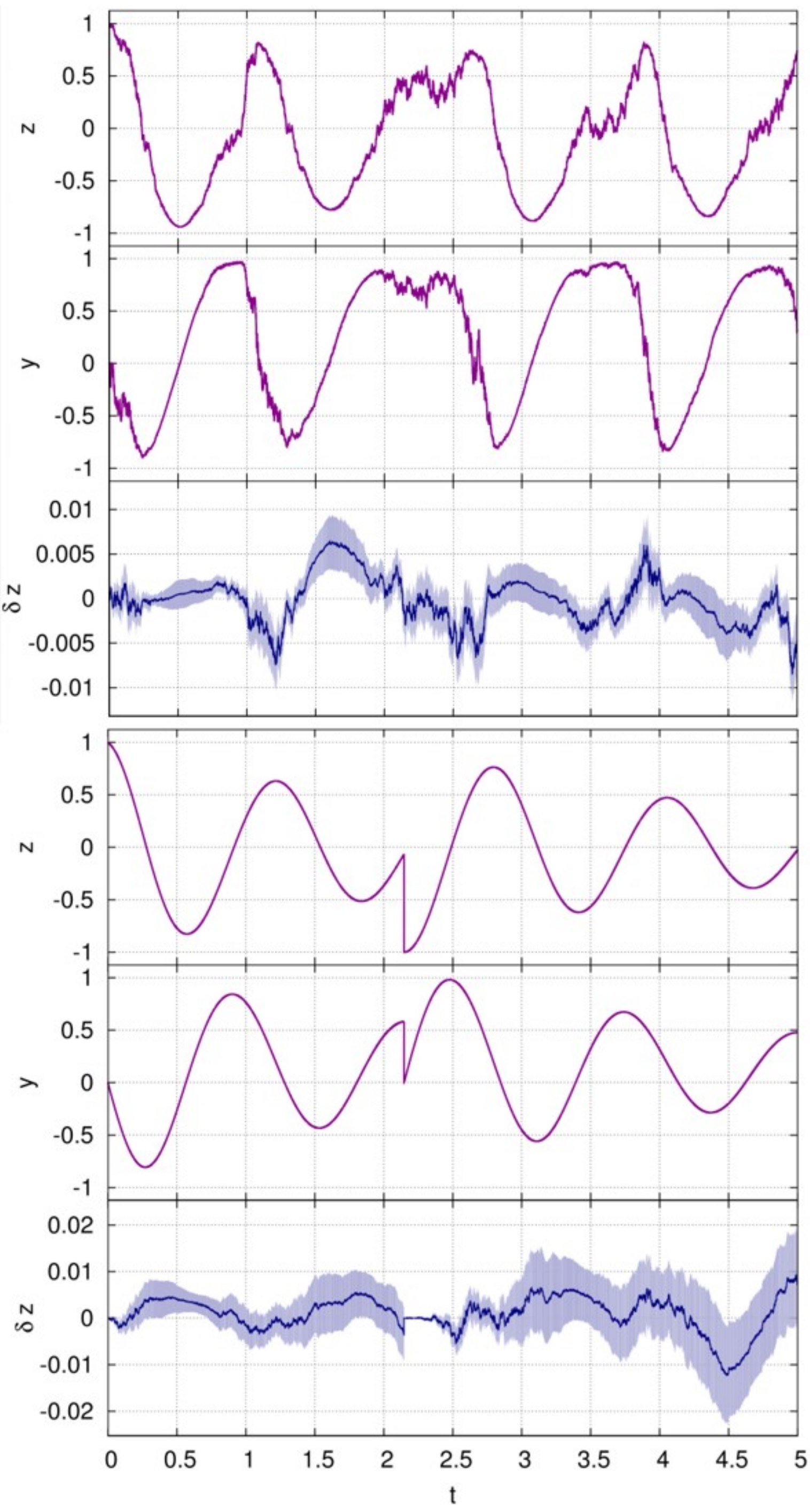}
\caption{[Color online] Comparison between the filtered state (purple) \unexpanded{$\rho_{\past{\bf  O}_{t}}$} and the ensemble average over $20000$ realizations of the multichannel quantum trajectory (blue) \unexpanded{$E_{\past{\bf  U}_{t}}[\rho_{\past{\bf  O}_{t},\past{\bf  U}_{t}}$]}. The curves are so close that no deviation is visible, so the deviation of results in the $z$ component of the Bloch Sphere (\unexpanded{$\delta z$}), together with error bars (pale blue) are also plotted.   The dynamical parameters are driving \unexpanded{$\Omega=5$},  phase \unexpanded{$\Phi=\pi/2$}, and total damping rate \unexpanded{$\Upsilon =\gamma+\Gamma = 1 $} with \unexpanded{$\gamma=0.5$}, \unexpanded{$\Gamma=0.5$}. The top three plots pertain to
unobserved record results  from  photodetections \unexpanded{$\past{\bf  U}_{t}=\past{ N}_{t}$} and  observed record result from homodyne detection  \unexpanded{$\past{\bf  O}_{t}=\past{Y}_{t}$}, while the bottom three to unobserved record results  from homodyne detection,  \unexpanded{$\past{\bf  U}_{t}=\past{Y}_{t}$} and observed record result from photodetections \unexpanded{$\past{\bf  O}_{t}=\past{ N}_{t}$.} 
}
\label{fig:rho0}
\end{center}
\end{figure}
The simulation of the quantum state smoothing has three stages. First, we simulate an initial realisation of the `true' state and the correspondent records $\both{\bf  O}_{\rm T}$ and $\both{\bf  U}_{\rm T}$. We do this by using the nonlinear CPQT  to generate the normalised state $\rho_{\rm T}$ (pure) from Eq.~\eqref{eq:multichanelrho}. $\both{\bf O}_{\rm T}$ and $\both{\bf U}_{\rm T}$ are drawn at random from the actual probability distribution for such records (Section \ref{sec:S2SMop}).  This will be used to illustrate the accuracy of the smoothed state as an estimate of the true state, and other features of their relations, in typical cases. Second, with the generated record $\both{\bf  O}=\both{\bf  O}_{\rm T}$ we assume the point of view of the observer Alice and numerically calculate a (mixed) filtered state $\rho_{\rm F}(t)=\rho_{\past{O}_t}(t)$, by the method detailed in Sec.~\ref{ssec:numcheck} below. This serves three purposes: 
it serves as a check on the accuracy of our CPQT method; for the case of unobserved jumps it is used to help compute the smoothed quantum state; and we contrast its behaviour with that of the smoothed state. Third, we calculate the smoothed state as follows. We generate a large ensemble of random measurement records for $\both{\bf  U}$ using a time-local ostensible probability distribution  $\wp_{\text{ost}}({\bf  u}_t)$, and for each record we calculate the associated unnormalized pure state $\tilde{\rho}_{\past{{\bf  U}_{t}}\past{{\bf  O}_{t}}}(t)$ (see Sec.~\ref{ssec:simun} below). It is this ensemble which allows checking the accuracy of our method against the filtered state (Sec.~\ref{ssec:numcheck}).  Finally, from $\both{\bf  O}_{\rm T}$ we calculate the effect operator  $\hat{E}_{\fut{\bf  O}_{t}}(t)$ using Eq.~\eqref{eq:Eretro} and with it obtain $\rho\sm(t)$ using  Eq.~\eqref{eq:tildap} and renormalising.

\subsection{ Simulations of the unnormalised possible true states}\label{ssec:simun}
To simulate the unnormalised state $\tilde{\rho}_{\past{{\bf  U}_{t}}\past{{\bf  O}_{t}}}(t)$ we start from $\rho_V$ in Eq.~\eqref{eq:rhoV}, then simulate the observed process without normalisation,
\begin{equation}\label{eq:rhoo}
\tilde{\rho}'_{{  o}_{t}}={\hat{M}_{{  o}_{t}}\rho_V \hat{M}_{{{o}_{t}}}^{\dagger}}.
\end{equation}
Next the effect of the hypothetical observations (${\bf  U}$) is implemented similarly, depending on the assumed nature of such process. 

For the case that the unobserved record is a quantum jumps process, we simulated it as a $\Delta n_t$ obtained from a Bernoulli distribution with ostensible probability $\wp_{\rm ost}(n_t:=1)=\Tr[{\hat{c}\,\rho_{{o}_t}(t)\,\hat{c}^{\dagger}}]\Delta t$, and the update 
\begin{equation}\label{eq:unrhony}\begin{array}{rl}
\tilde{\rho}_{{n_t,{o}_t}}( t+\Delta t)=&\Delta n_t\hat{M}_1\tilde{\rho}'_{{o}_t}\hat{M}_1^{\dagger}+(1-\Delta n_t){\hat{M}_0\tilde{\rho}'_{{o}_t}\hat{M}_0^{\dagger}},
\end{array}\end{equation}
with $\hat{M}_{n_t}$ following Eq.~\eqref{eq:measopdt}. Note that the ostensible probability we use is {\em not} the same at each time step, 
as it it depends on the observed record $\both{O}$. Nevertheless, given that record, the ostensible distribution does factorize for different times, 
as stated earlier. 

For the case that the unobserved record is a quantum diffusive process, we generate the $\dd y_t$ as a random Gaussian variable following an ostensible probability distribution with mean zero and variance $1/\Delta t$, as per Sec.~\ref{Sec:4B}. The corresponding map is 
\begin{equation}\label{eq:unmultichanelrho}
\tilde{\rho}'_{{y_t},{o}_t}= {\hat{M}_{y_t}\tilde{\rho}'_{{o}_t}\hat{M}_{y_t}^{\dagger}},
\end{equation}
with $\hat{M}_{y_t}$ following Eq.~\eqref{eq:MytCPQT}. Rather than the Gaussian used here, it would have been possible, similarly to the jumps case, to use a Guassian with statistical moments given by Eqs.~\eqref{eq:mu} and \eqref{eq:sigma} respectively, with expectations $\braket{\bullet}$ calculated using  ${\rho_{\past{\bf  O}_{t}}(t)}$. 
 This may have been slightly more efficient in terms of required ensemble size (see below) but, as mentioned in Section \ref{sec:QTF}, the resultant quantum trajectories are independent of the choice of the ostensible probabilities.

\subsection{Filtering, Retro-filtering and Numerical checks}\label{ssec:numcheck}

From the generated observed record $\both{\bf  O}$ we calculate Alice's filtered state using the map
\begin{equation}
\rho_{\past{\bf  O}_{t}} (t+\Delta t)  =\frac{\hat{M}_{{ o}_{t}}\hat{V}{\rho}( t+\Delta t) \hat{V}\dg\ \hat{M}_{{ o}_{t}}^{\dagger}}{\Tr[\hat{M}_{{ o}_{t}}\hat{V}{\rho}(t+ \Delta t)\hat{V}\hat{M}_{{ o}_{t}}^{\dagger}]}.
\end{equation}
Here ${\rho}( t+\Delta t)$ is what results from averaging over the unobserved results, according to Eqs.~\eqref{eq:mecpqtjumps} or \eqref{eq:mecpqtdiff},  depending on which channel is unobserved by Alice, with $\rho(t)$ in those equations being $\rho_{\past{\bf  O}_{t-\Delta t}}(t)$.  It is important to note that this filtered state must coincide with the ensemble average that can be obtained from the `true' state $\rho_{\past{\bf  O}_{t}}=E_{\past{\bf  U}_{t}}[\rho_{\past{\bf  O}_{t},\past{\bf  U}_{t}}]$, and such relation can be used to test the accuracy of the simulations,  as we now explore.

 We   remark \blk that it is necessary to use unnormalized states in such an average, even for filtering, as explained in detail in Ref.~\cite{G&W05}.  {T}he graphs in Fig.~\ref{fig:rho0}, show the state matrix components of the original filtered state $\rho_{\past{\bf  O}_{t}}$ and the one obtained after averaging $\rho_{\past{\bf  O}_{t},\past{\bf  U}_{t}}$ over an ensemble of  20000  smoothed unobserved records $\past{\bf  U}_{t}$.  We also present the deviation from the original filtered state $z$ component.  The deviation is seen to be consistent with zero, given the size of the errors, which are very small.  That is, we can confirm that the average over the unobserved records, with the correct weighting probabilities, does give back the filtered state with reasonable uncertainty. These results provide us with enough support to be confident that the smoothed state calculation is also correct, although there is not a direct way to verify it. Thus we can now be confident that any increase in purity in $\rho\sm (t) $ over $\rho\fil(t)$, for example, 
 is reliable and only due to the quantum state smoothing method. \blk

\begin{figure}[!htbp]
\begin{center}
\includegraphics[width=0.6\columnwidth]{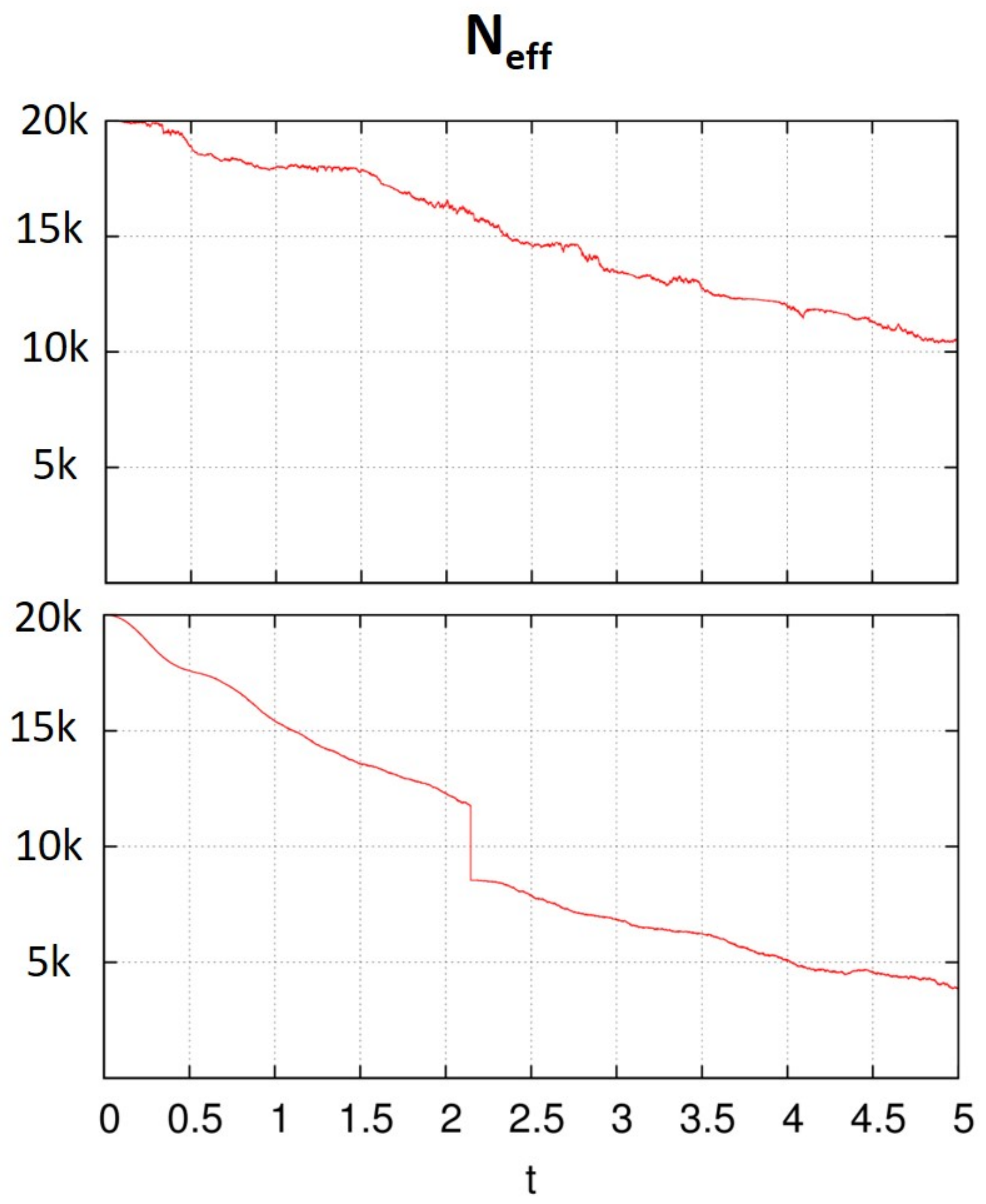}
\caption{[Color online]  Effective number of trajectories  \unexpanded{$N_{ \rm eff}$} in the ensemble average over $20000$ realizations of the multichannel quantum trajectory (blue) \unexpanded{$E_{\past{\bf  U}_{t}}[\rho_{\past{\bf  O}_{t},\past{\bf  U}_{t}}$]}. The dynamical parameters are driving \unexpanded{$\Omega=5$},  phase \unexpanded{$\Phi=\pi/2$}, and total damping rate \unexpanded{$\Upsilon =\gamma+\Gamma = 1 $} with \unexpanded{$\gamma=0.5$}, \unexpanded{$\Gamma=0.5$}. (Bottom) Unobserved record results  from  photodetections \unexpanded{$\past{\bf  U}_{t}=\past{ N}_{t}$} and  observed record result from homodyne detection  \unexpanded{$\past{\bf  O}_{t}=\past{Y}_{t}$}. (Top) Unobserved record results  from homodyne detection,  \unexpanded{$\past{\bf  U}_{t}=\past{Y}_{t}$} and observed record result from photodetections \unexpanded{$\past{\bf  O}_{t}=\past{ N}_{t}$}. }
\label{fig:Neff}
\end{center}
\end{figure}

We wish to briefly draw attention to the fact the size of the errors in Fig.~\ref{fig:rho0} noticeably increase in time with the bottom case, where  $\past{\bf  U}_{t}=\past{Y}_{t}$. To explain this difference, we show in Fig.~\ref{fig:Neff}  the effective number of trajectories 
\beq N_{ \rm eff}=\frac{\left(\sum_{\past{\bf  U}_{t}}\wp_{\text{ost}}(\past{\bf  U}_{t})\,{\Tr[\tilde{\rho}_{\past{\bf  O}_{t},\past{\bf  U}_{t}}(t)]}\right)^2}{\sum_{\past{\bf  U}_{t}}\wp_{\text{ost}}(\past{\bf  U}_{t})\,{\left(\Tr[\tilde{\rho}_{\past{\bf  O}_{t},\past{\bf  U}_{t}}(t)]\right)^2}},\eeq in the ensemble average of unobserved records for both cases considered. The evolution of $N_{ \rm eff}$ shows that in the unobserved jumps case, the effective number of trajectories decays faster than in the unobserved diffusion case.  The standard deviation of the mean is inversely  proportional to    $\sqrt{N_{ \rm eff}}$, so this is consistent with the  more rapidly growing uncertainties for the latter case. 

\begin{figure}[!htbp]
\begin{center}
\includegraphics[width=0.8\columnwidth]{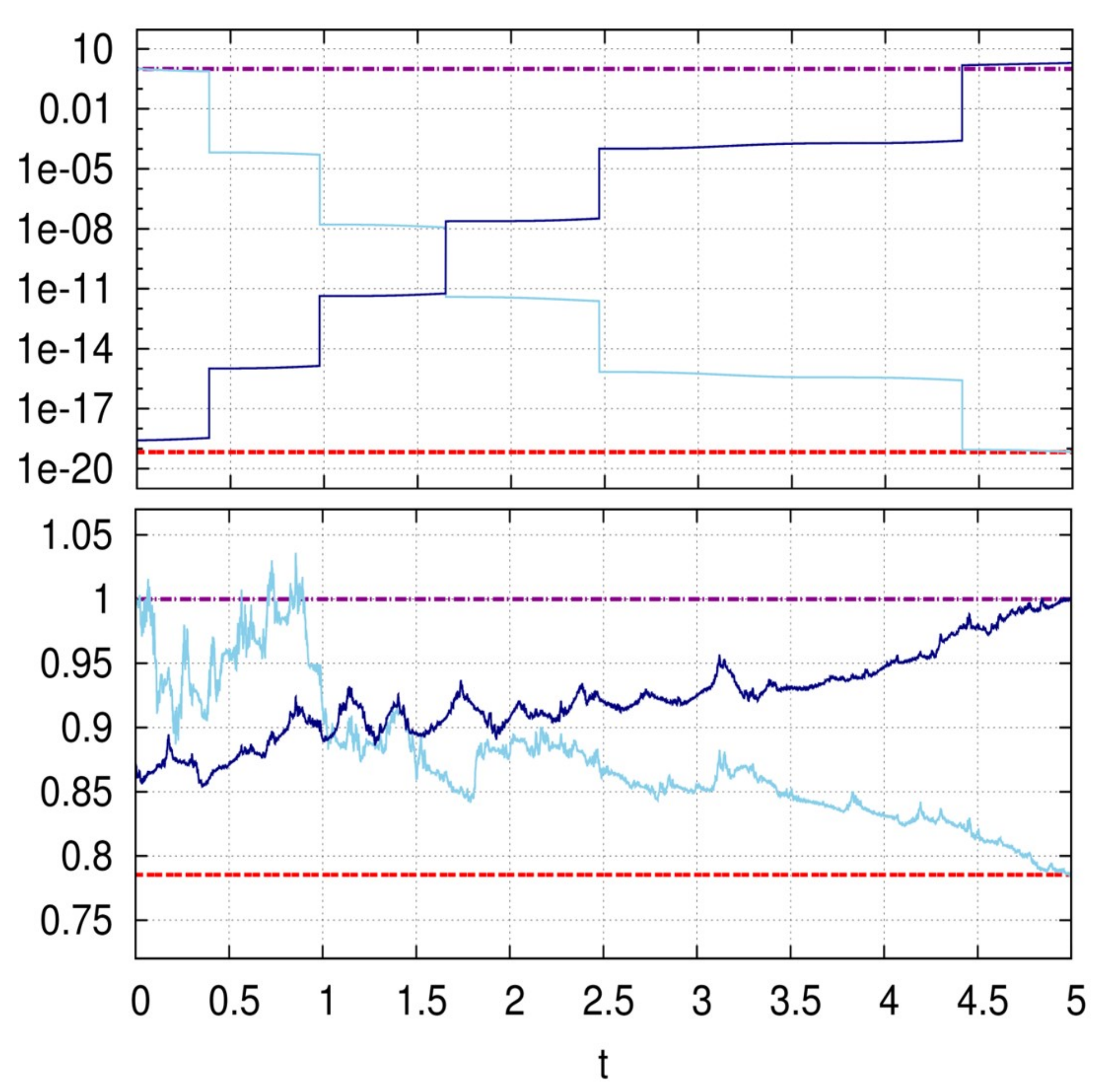}
\caption{[Color online] Effect operator test. These figures test the evolution of constants \unexpanded{$\Tr[{\tilde\rho}_{{\past{\bf  O}}_{t}}\, {\hat {E}}_{{\fut{\bf  O}}_{t}}]=\Tr[{\tilde\rho}_{{\past{\bf  O}}_{T}}]$} (dotted red line)  and \unexpanded{$\Tr[{\rho}_{{\past{\bf  O}}_{t}}]$=1} (dot dashed purple line) . The evolutions of \unexpanded{$\Tr[{\tilde\rho}_{{\past{\bf  O}}_{t}}]$} (irregular light blue line) and \unexpanded{$\Tr[ {\hat {E}}_{{\fut{\bf  O}}_{t}}]$} (irregular dark blue line) are also presented. The dynamical parameters are $\Omega = 5$ and total damping rate \unexpanded{$\Upsilon =\gamma+ \Gamma=1$}. (Top) The observed records result  from homodyne detection  \unexpanded{$\past{\bf  O}_{t}=\past{ Y}_{t}$} with  \unexpanded{$\gamma=1, \Gamma=0$} and phase $\Phi = \pi/2$.  (Bottom) The observed records result  from photodetections  \unexpanded{$\past{\bf  O}_{t}=\past{ N}_{t}$} with  \unexpanded{$\gamma=0, \Gamma=1$}. }
\label{fig:TestEy}
\end{center}
\end{figure}

The smoothing estimation depends as much on the retrofiltering part as it does on the filtering one.  In this case the \emph{effect operator} from Eq.~\eqref{eq:Eretro}  evolves backwards from the final uninformative effect $\hat{E}(T)=I$ towards $\hat E\retro(t) \equiv \hat E_{\fut{\bf  O}_{t}}(t)$,  conditioned on the record $\fut{\bf  O}_{t}$ in the future of $t$. This condition determines the consistency between retrofiltering and filtering and is vital for quantum state smoothing. We use the completely positive trajectories in the retrofiltering step of simulation using the  adjoint map ${\cal M}_{{\bf  o}_{t}}^{\dagger}$ to guarantee the regularity in the evolution.  In Fig.~\ref{fig:TestEy} we show the evolution in time of $\Tr[{\tilde{\rho}}_{{\past{\bf  O}}_{t}}\, {\hat {E}}_{{\fut{\bf  O}}_{t}}]$ verifying that it is a constant value, $\Tr[{\tilde{\rho}}_{{\past{\bf  O}}_{T}}]$, determined by the unnormalised filtered state at the end of the interval $t=T$. This test allows us to confirm that the effect operator and the unnormalised states are consistent and reliable to calculate the quantum state smoothing. The graph also shows the evolution of the normalised state trace, which as expected is one at all times.

\subsection{Results for smoothing calculation}
\begin{figure}[!bp]
     \begin{center}
            \includegraphics[width=0.5\textwidth]{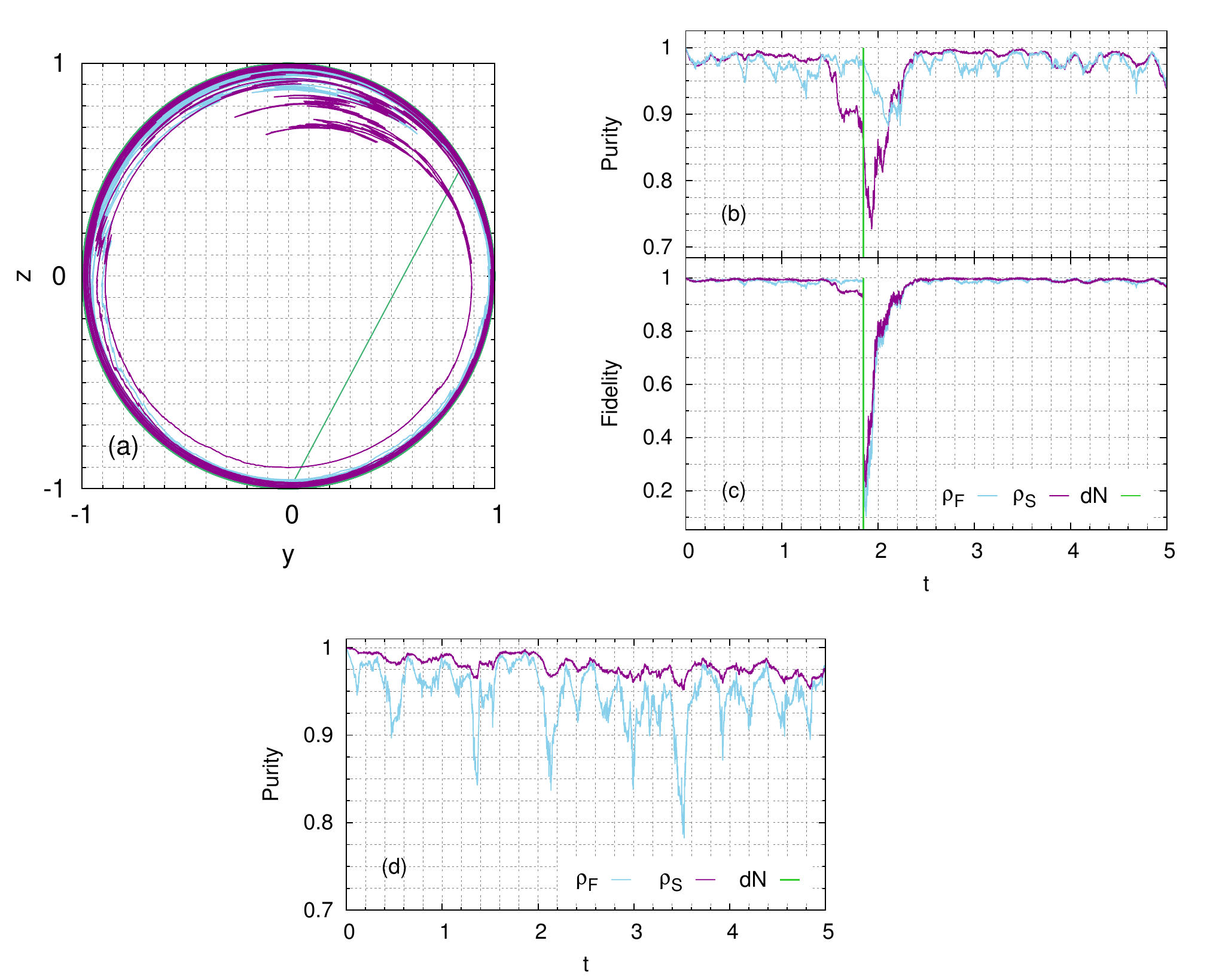} 
     \vspace{-0.5cm}  
    \end{center} 
    \caption{[Color online]  (a) Trajectories in the Bloch sphere  for our model system with $\Omega = 20$, $\Upsilon=1$, $\eta = 10/11$, $\Gamma=10\gamma$ and $\Phi=\pi/2$.  The states shown are $\rho\fil$ (filtered, blue), $\rho\sm$ (smoothed, purple) and $\rho\god$ (`true', green) for a case where the `true' record includes a jump. We also plot the purities (b) and fidelities with $\rho\god$  (c) of these $\rho\fil$ and $\rho\sm$. The purities for a record with no jump are shown in (d).  To compute $\rho_{\sm}$ we average over an ensemble of $10^4$ hypothetical unobserved records \unexpanded{$\both{N}$}. Reproduced from \cite{G&W15}.}
   \label{fig:singletraj}
\end{figure}

 Having established that all the elements are in place for an accurate calculation of quantum state smoothing, we turn now to the results previously obtained by our method in Ref.~\cite{G&W15}. There we showed that, as expected, quantum state smoothing gives, on average (over the observed and unobserved records), a better (more faithful) estimate of the true state than does quantum state filtering. We did this taking  the observed record to be a homodyne photocurrent ($\both{Y}$) and the unobserved record to be photodections ($\both{N}$). We considered both  Y-homodyne ($\Phi=\pi/2$)  and  X-homodyne ($\Phi=0$), with the improvement in average fidelity offered by smoothing being markedly better in the former case. We do not reproduce those results here, but rather turn to the typical trajectories for the quantum smoothed state. These were shown in Ref.~\cite{G&W15} for the case $\Phi=\pi/2$, for the parameters given in the caption of Fig.~\ref{fig:singletraj}.

 Specifically, in Ref.~\cite{G&W15}, we showed two trajectories, as reproduced in Fig.~\ref{fig:singletraj}. One was for the more likely case where there were no photodections in the record $\both{N}$, and the other for the less likely case where there was a photodection in the unobserved record, roughly in the middle of the total time period $[0,T]$ where $T=5/\Upsilon$. That is, in the latter case, the true state $\rho\true(t) = \rho_{\past{\nbf Y_{t}},\past{\nbf N_{t}}}(t)$ changed discontinuously at a particular time $t_j$ arising from the simulation. In the former case, the smoothed state was consistently more pure than the filtered state, as expected. In the latter case, surprisingly, the purity of the smoothed state was markedly lower than that of the filtered state in the region of $t_j$. Most noticeably, the change in the purity of $\rho\sm$ {\em anticipated} the jump in $\rho\true$,  which it can do because it uses observed information from the future of that jump. We interpreted the lower purity (compared to the non-anticipating $\rho\fil$) as being due to the smoothing algorithm's being uncertain about the precise timing of the jump. Similarly, the fidelity of $\rho\sm$ to $\rho\true$ was observed to decrease below that of $\rho\fil$ prior to the jump, but to be higher after the jump. 

\begin{figure}[!btp]	
\begin{center}
\includegraphics[width=0.9\columnwidth]{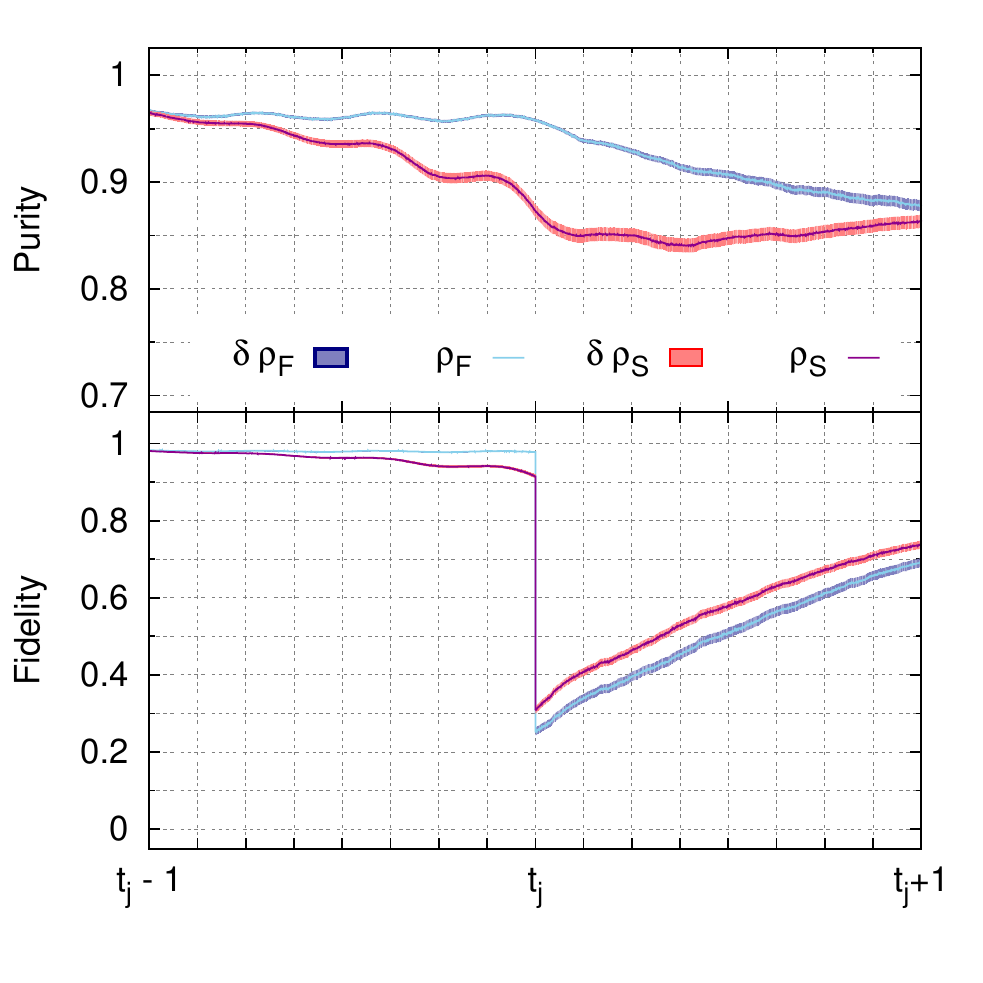}
\caption{[Color online]  Average purity and fidelity around a single jump for the case considered for filtered (]blue line) and smoothed states (red line) with their corresponding statistical error $\delta \rho_C$. The average has been calculated with the 530 trajectories with one jump in the interval $[2,4]$ out of a total interval $[0,5]$, from the  $ 5000 $ 
observed and unobserved record pairs \blk Each smoothed trajectory is calculated with $10^4$ estimated \unexpanded{$\both{N}$} records. The dynamical parameters are $\Omega = 20$, $\Upsilon=1$, $\eta = 10/11$, $\Gamma=10\gamma$ and $\Phi=\pi/2$, the same parameters used in Ref.~\cite{G&W15}.}
\label{fig:atjump}
\end{center}
\end{figure}

The question naturally arises as to whether the remarkable behaviour for the estimated states seen for the case of a single unobserved jump in Ref.~\cite{G&W15} can be proven to be typical.  Recall that the estimation of quantum state filtering and smoothing are generated purely from the observed record. We can address the typicality question by simulating many possible observed and unobserved records simultaneously as in Sec.~\ref{sec:S2SMop}, and select only those with one unobserved jump neither near the beginning nor the end of the smoothing interval. Here, as a direct application of the completely positive quantum trajectories,  and the quantum state smoothing theory presented in Sec.~\ref{Sec:QSS}, we  present in Fig.~\ref{fig:atjump} 
the ensemble average of purity and fidelity around the time where a jump occurs. We generated 5000 observed and unobserved record pairs over the  total interval $[0,5]$, under the same conditions previously described for one trajectory, and calculated averages using the 530 trajectories with one unobserved jump in the interval $[2,4]$. (Here we are using $\Upsilon^{-1}$ as the time unit.)  The results indicate that the capacity of anticipating the occurrence of an unobserved jump is not unusual. The purity decays previous to the jump  for $\rho\sm$ but does not do so for $\rho\fil$, as expected.  This is also seen in the fidelity, and, again as seen with a single trajectory, the fidelity for $\rho\sm$ is higher than for $\rho\fil$ after the jump.

\section{Conclusions}
We have introduced an extension to higher orders in $\Delta t$ of the quantum trajectories formalism, that can guarantee the complete positivity of the maps in simulations.  This extension has been done for both quantum jumps and quantum diffusive trajectories and for each of them, the completeness relation has been evaluated and the corresponding modified master equation has been obtained to verify the complete positivity of the new maps. We applied the method to quantum state smoothing, a recently proposed application of the quantum trajectories formalism. We demonstrated the usefulness of our method for precise simulations. Specifically our method  allows a fair analysis of the advantages of quantum state smoothing compared with the standard quantum filtering. This was useful in enabling us to explore an interesting regime for comparison where an event (a quantum jump) occurs but is unseen by the observer performing the filtering and smoothing. 
\acknowledgments We want to thank Michael Hall and Joshua Combes for inspiring and enlightening discussions. This research was supported by the Australian Research Centre of Excellence Council Program CE170100012. We acknowledge the traditional owners of the land on which this work was undertaken at Griffith University, the Yuggera people.

\bibliography{references,books}

\end{document}